\documentclass[prc,superscriptaddress,showpacs,twocolumn]{revtex4}
\usepackage{graphicx,dcolumn,array,bm,amsmath}

\newcommand{\re}{\mathrm{Re~}}
\newcommand{\im}{\mathrm{Im~}}
\newcommand{\cJ}{\mathcal{J}}
\renewcommand{\vec}{\boldsymbol}
\newcommand{\Tr}{\mbox{Tr}}

\begin{document}

\title{\boldmath{Search for the Skyrme-Hartree-Fock Solutions for Chiral
Rotation in $N$=75 Isotones}\unboldmath}
\author{P. Olbratowski}
\email[]{Przemyslaw.Olbratowski@fuw.edu.pl}
\affiliation{Institute of Theoretical Physics, Warsaw University,
             ul. Ho\.za 69, PL-00681, Warsaw, Poland}
\author{J. Dobaczewski}
\email[]{Jacek.Dobaczewski@fuw.edu.pl}
\affiliation{Institute of Theoretical Physics, Warsaw University,
             ul. Ho\.za 69, PL-00681, Warsaw, Poland}
\author{J. Dudek}
\email[]{Jerzy.Dudek@ires.in2p3.fr}
\affiliation{\it Institut de Recherches Subatomiques
                 IN$_2$P$_3$-CNRS/Universit\'e Louis Pasteur,
                 F-67037 Strasbourg Cedex 2, France}

\begin{abstract}

A search for the self-consistent solutions for the chiral rotational
bands in the $N=75$ isotones, $^{130}$Cs, $^{132}$La, $^{134}$Pr and
$^{136}$Pm is performed within the Skyrme-Hartree-Fock
cranking approach using SKM$^*$ and SLy4 parametrizations. The
dependence of the solutions on the  time-odd contributions in the
energy functional is studied. From among the considered four
isotones, self-consistent chiral solutions are obtained only in
$^{132}$La. The microscopic calculations are compared with the
$^{132}$La experimental data and with results of a classical
model that contains all the mechanisms underlying the chirality of
the collective rotational motion. Strong similarities between the HF
and classical model results are found. The suggestion formulated
earlier by the authors that the chiral rotation cannot exist below a
certain critical frequency is further illustrated and discussed,
together with the microscopic origin of a transition from the
planar to chiral rotation in nuclei. We also formulate
the separability rule by which the Tilted-Axis-Cranking solutions
can be inferred from three independent Principal-Axis-Cranking
solutions corresponding to three different axes of rotation.

\end{abstract}

\pacs{21.30.Fe, 21.60.Ev, 21.60.Jz}

\maketitle

\section{Introduction}

Since the original 1997 work of Frauendorf and Meng \cite{Fra97a}, the
phenomenon of chiral rotation in atomic nuclei attracts quite a
significant attention. The effect is expected to occur in nuclei having
stable triaxial deformation, and in which there are a few high-$j$
valence particles and a few high-$j$ valence holes. The former drive
the nucleus towards prolate, and the latter towards oblate shapes.
The interplay of these opposite tendencies may favor a stable triaxial
deformation. For such a shape, the valence particles and holes align
their angular momenta along the short and long axes of the density
distribution, respectively. However, the nuclear-bulk moment of inertia
with respect to the medium axis is the largest, which favors
collective rotation about that axis. Thus, the particle, hole, and
collective angular momentum vectors are aplanar, and may form either
a left-handed or a right-handed set. In this way, the two
enantiomeric forms may give rise to pairs of rotational bands, which
are called chiral doublets.
It is expected that the energy splitting between the partners in such
doublets is very small, and in fact the authors of Ref.\ \cite{Sta01b}, who have
analyzed the experimental results, have used essentially the argument
'by elimination' - the bands were suggested to be chiral partners mainly
because their properties could not be explained within
other scenarios used by the authors.

The first doublet band, later reinterpreted as chiral \cite{Fra97a},
was found in 1996 by Petrache {\it et al.} in $^{134}$Pr
\cite{Pet96a}. Now, about 15 candidate chiral doublet bands are known in
the $A\approx130$ region, and about 10 in the $A\approx100$ region.
The bands in the $A\approx130$ nuclei are assigned to the simplest
chiral configuration, in which there is one proton particle on the
$\pi h_{11/2}$ orbital, and one neutron hole on the $\nu h_{11/2}$ orbital.
Configurations for $A\approx100$ nuclei usually involve one $\pi g_{9/2}$ proton
hole and one $\nu h_{11/2}$ neutron particle orbitals. A few cases with more than
one active particle or hole were also found \cite{Zhu03a}. So far,
experimental information about absolute B(E2) and B(M1) values for
transitions within the observed bands is available only for
$^{128}$Cs and $^{132}$La, from recent lifetime measurements by
Grodner \cite{Gro04a,Gro05a} and Srebrny \cite{Sre05a}, and
collaborators.

On the theoretical side, chiral rotation has been extensively studied
by using various versions of the Particle-Rotor Model (PRM), in which
the nucleus is represented by the valence particles and holes coupled
via the quadrupole-quadrupole interaction to a rotator, often
described within the Davydov-Fillipov model \cite{Dav58a} with moments of
inertia given by the irrotational-flow formula \cite{Boh75a}, see, e.g.,
Refs.\ \cite{Sta02a,Koi03a,Pen03a}.
However, the main concept of rotational chirality summarized above,
came from considerations within the Frauendorf's mean-field
Tilted-Axis-Cranking (TAC) model \cite{Fra93a}, which is used in
parallel with the PRM. That model is a straightforward generalization
of the standard cranking approach to situations where the axis of
rotation does not coincide with any principal axis of the mass
distribution.

Up to now, all the TAC calculations for chiral
rotation (see Refs.\ \cite{Dim00a,Hec01a,Sta01b,Rai03a,Zhu03a} for
examples) were performed within a phenomenological approximation, in
which the mean field is given by a simple model potential. A more
fundamental description requires self-consistent methods, which would
provide a strong test of the stability of the proposed chiral
configurations with respect to the core degrees of freedom.
Self-consistent methods are also necessary to take into account all
kinds of polarization of the core by the valence particles and full
minimization of the underlying energies with respect to all
deformation degrees of freedom, including deformations of the current
and spin distributions. Application of self-consistent methods to the
description of chiral rotation is the subject of the present paper.

Our study concerns four $N=75$ isotones, $^{130}$Cs, $^{132}$La,
$^{134}$Pr, and $^{136}$Pm, which are the first nuclei in which
candidate chiral bands were systematically studied \cite{Sta01b}. We
used the Hartree-Fock (HF) method with the Skyrme effective
interaction. The results were obtained for two Skyrme parameter sets,
SLy4 \cite{Cha97a} and SKM* \cite{Bar82a}, and the role of terms in
the mean field that are odd under the time reversal was examined.
Calculations were carried out by using a new version of the code HFODD (v2.05c)
\cite{Dob00a,Dob04a,Dob05a}, which was specially constructed by the
authors for the purpose of the present study. From among the
considered four isotones, self-consistent chiral solutions were
obtained in $^{132}$La. A brief report on the results obtained in
$^{132}$La was given in Ref.\ \cite{Olb04a}.

The paper is organized as follows. In Section~\ref{tac_sec} we
discuss some characteristic aspects of the TAC calculations within the
self-consistent framework. Section~\ref{previous_sub} briefly recalls
previous studies on chiral rotation in the concerned isotones, and
Section~\ref{detail_sub}
describes all technical details of the present calculations -- in
particular the way in which the role of time-odd nucleonic densities
was examined. Energy minima obtained for non-rotating states are
listed in Section~\ref{enemin_sub}. In Section~\ref{pac_sub},
rotational properties of the valence nucleons and of the core are
examined within standard Principal-Axis Cranking (PAC). In
Section~\ref{clasic_sub} we solve a simple classical model of chiral
rotation and show that such a rotation cannot exist below a certain critical
angular frequency. The HF solutions for planar and chiral rotation are
presented in Sections~\ref{planar_sub} and \ref{chiral_sub},
respectively. In Section~\ref{separa_sub} we demonstrate that our results
obtained for the
three-dimensional rotation can actually be represented as a sum of
three independent one-dimensional rotations about the principal axes.
The values
obtained for the critical frequency and the agreement of our results
with experimental level energies are discussed in
Section~\ref{discus_sec}. In Appendix~\ref{align_app}, we study
response of the single-particle (s.p.) angular momenta to
three-dimensional rotation, and
introduce the notions of \emph{soft} and \emph{stiff} alignments.

\section{Hartree-Fock Tilted-Axis-Cranking calculations}
\label{tac_sec}

So far, the TAC model has been described in
the literature only in its phenomenological variant \cite{Fra93a}.
Therefore, in this Section we give several details that are specific
for its self-consistent implementation. The discussion concerns
mainly the way of iteratively solving the HF equations which is
adopted in this work.

As far as non-rotating states are concerned, the HF method consists
in minimizing the expectation value of the many-body Hamiltonian,
$\hat{H}=\hat{T}+\hat{V}$, in the trial class of Slater determinants.
Here, $\hat{T}$ is the kinetic-energy operator, and $\hat{V}$ is a
two-body effective interaction. Equivalently, one can formulate the
method in terms of the energy density functional $E\{\rho\}$, which is minimized
with respect to the one-body density matrix $\rho$ on which it depends, and
this latter representation is used in the present study. Since
$\hat{H}$ or the density functional $E\{\rho\}$ are invariant under rotations in
space, it is clear that the HF solution is defined only up to an
arbitrary rotation. For each solution it is useful to introduce an
intrinsic frame of reference, whose axes we define as principal axes
of the tensor of the electric quadrupole moment of the mass
distribution. Due to the mentioned arbitrariness, this frame can be
rotated with respect to the frame originally used to solve the HF
equations, which we refer to as the program (or computer-code) frame. The program frame
is the one defined by the axes $x$, $y$ or $z$, used for solving the
mean-field equations, e.g., in a computer code \footnote{It would be
unfortunate to call it laboratory frame, because in the common
interpretation of the cranking model, both the program and the
intrinsic frames rotate in the laboratory system. On the other hand,
the program and intrinsic frames do not move with respect to each
other -- they differ only by the orientation of their axes.}.

To describe rotational excitations within the TAC approach, in the
program frame one imposes a linear constraint on angular momentum
and minimizes the expectation value of the Routhian,
\begin{equation}
\label{HHwJ_cranking_eqn}
\hat{H}'=\hat{H}-\vec{\omega}\cdot\hat{\vec{I}}~,
\end{equation}
or the energy density in the rotating frame,
\begin{equation}
\label{EwJ_cranking_eqn}
E'\{\rho\}=E\{\rho\}-\vec{\omega}\cdot\Tr(\hat{\vec{I}}\rho)~,
\end{equation}
where $\hat{\vec{I}}$ is the total angular momentum operator, and
vector $\vec{\omega}$ is composed of three Lagrange multipliers; it is called rotational
frequency vector. Its components in the program frame are fixed as part of
the definition of $\hat{H}'$ or $E'\{\rho\}$. Because of the
rotational invariance, solutions obtained for the same length, but
different directions of $\vec{\omega}$ differ only by their
orientation in the program frame, so that only the length,
$\omega=|\vec{\omega}|$, has physical meaning.

Within the HF procedure,
one obtains that the sought Slater determinant
or the one-body density are built of the eigenstates of the s.p.\
Routhian,
\begin{equation}
\label{hhwj_eqn}
\hat{h}'=\hat{h}-\vec{\omega}\cdot\hat{\vec{I}}~,
\end{equation}
where the mean-field Hamiltonian,
$\hat{h}=\hat{T}+\hat{\Gamma}=\delta{E\{\rho\}}/\delta\rho$, is a sum
of the kinetic energy, $\hat{T}$, and the mean field, $\hat{\Gamma}$. A phenomenological approximation to the HF approach consists in
replacing $\hat{\Gamma}$ with a model potential, like the Nilsson
potential \cite{Nil55a}, whose deformation dependence is usually parametrized
in terms of the multipole deformations of the
nuclear surface. Then, the expectation value of the Routhian,
$\hat{h}'$, is minimized over the deformation parameters of the potential
within the standard Strutinsky method \cite{Str67a,Str67b}.

Obviously, only the relative orientation of the angular momentum
vector with respect to the nuclear body carries physical
information. Common orientation of the angular momentum vector and
nucleonic densities with respect to the program frame
can thus be arbitrary. In the phenomenological method, where the
orientation in space of the nuclear surface is under control via the
multipole deformations $\alpha_{\lambda\mu}$, one can take advantage
of this fact and fix $\alpha_{21}=\im\alpha_{22}=0$ so that the
intrinsic and program frames coincide. Minimization of the
expectation value of the Routhian, $\hat{h}'$, at a given magnitude
of $\omega$ is then performed by varying the direction of
$\vec{\omega}$ and all $\alpha_{\lambda\mu}$ except of $\alpha_{21}$
and $\im\alpha_{22}$.

In the HF method, however, the Euler angles defining the orientation
of the intrinsic axes in the program frame are not free variational
parameters, but complicated functions of densities, which in turn
change from one HF iteration to another. The only possible
way to make the two frames coincide is by imposing dynamical
constraints on the off-diagonal components of the quadrupole tensor
and requiring that they vanish. Then, one can vary the cranking
frequency vector like in the phenomenological approach. This is
actually the only way to proceed if the energy dependence on the
intrinsic orientation of $\vec{\omega}$ is sought. However,
constraints strong enough to confine the nucleus easily lead to
divergencies, and adjusting their strengths properly may be a
cumbersome task. If only the energy minimum is of interest, a more natural
and incomparably faster way is to fix $\vec{\omega}$ in the program frame
and let the mean potential reorient and conform to it
self-consistently in the course of the iterations. The
intrinsic axes now do become tilted with respect to the program
frame.

The Kerman-Onishi theorem \cite{Ker81a} states that in each
self-consistent solution the total angular momentum vector,
$\vec{I}=\langle\hat{\vec{I}}\rangle$, is parallel to $\vec{\omega}$.
In calculations, the angle between those vectors converges
to zero very slowly in terms of the HF iterations, because the whole
nucleus must turn in the program frame in order to align its
$\vec{I}$ with the fixed $\vec{\omega}$. Therefore, a much faster
procedure is to explicitly reset $\vec{\omega}$ in each iteration to
make it parallel to the current $\vec{I}$, while keeping its length,
$\omega$, constant \cite{Dob04a}. This purely heuristic procedure
does not correspond to a minimization of any given Routhian. However,
once the self-consistent solution is found, it is the Routhian for
the final angular frequency that takes its minimum value.

Some quantities, like mean angular momenta and multipole moments,
are easiest to discuss only when expressed in the intrinsic frame
of the nucleus, but it is a natural way to calculate them first in
the program frame. Since the two frames do not necessarily coincide,
one has to find the axes of the intrinsic frame by diagonalization
of the quadrupole deformation tensor and to transform the considered
quantities into that frame by use of the Wigner matrices. (Such a
procedure may fail when the solutions have vanishing quadrupole moments, but
in the present paper such cases are of no interest and will not be discussed).

The HF TAC solutions are arbitrarily tilted with respect to the
program frame, and, moreover, their orientation is not known a
priori. Therefore, when solving the problem numerically one should
ensure such conditions that the same solution be represented equally
well in all orientations. In particular, the energy must not depend
on the orientation. If the s.p.\ wave-functions are expanded onto a
basis, this means that the choice of the s.p.\ basis and of the
basis cut-off must not privilege
any axis of the reference frame. In the case of the Cartesian harmonic-oscillator (HO) basis used in the
present study, this by definition amounts to taking the three oscillator
frequencies equal and including only the entire HO shells. To obtain a
reasonable description of deformed nuclei in such non-deformed bases,
the only way is to use sufficiently many HO shells.


\section{Results}

\subsection{Experiment and previous calculations}
\label{previous_sub}

\begin{figure*}
\begin{center}
\includegraphics{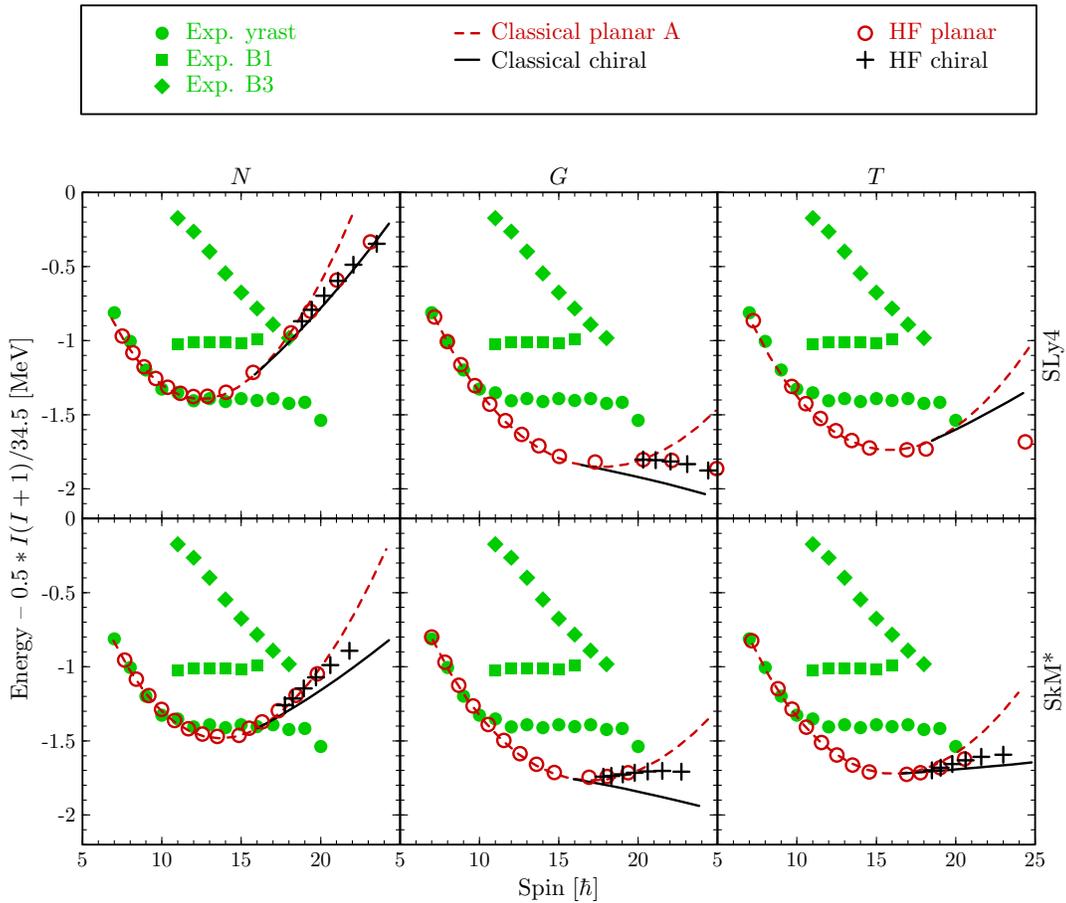}
\end{center}
\caption{(color online). Energies obtained from the HF TAC
         calculations, open circles and plus symbols, and from the classical model
         presented in Sec.\ III.D, dashed and solid lines, compared with
         the experimental data on the candidate chiral partner bands in
         $^{132}$La, full symbols. The HF results obtained for the SLy4 and SkM*
         forces are shown for the $N$, $G$, and $T$ time-odd fields included
         (see text).}
\label{laener_fig}
\end{figure*}

It the present study, the HF solutions corresponding to chiral rotation
were sought in four $N=75$ isotones, $^{130}$Cs, $^{132}$La,
$^{134}$Pr, and $^{136}$Pm. These are the first nuclei in which
candidate chiral bands were observed and systematically studied
\cite{Sta01b}. For the most recent experimental data refer to
\cite{Koi03a} for $^{130}$Cs, to \cite{Sta02a,Tim03a,Sre05a} for
$^{132}$La, to \cite{Sta01b,Rob03a} for $^{134}$Pr, and to
\cite{Har01a} for $^{136}$Pm. Absolute values of spins and
reduced transition probabilities were measured only in $^{132}$La.
The possible chiral bands known in this nucleus are shown in
Fig.~\ref{laener_fig} with full symbols. The lowest one
(circles) is the yrast in positive parity. Two closely-lying excited bands, B1
and B3 (squares and diamonds) are known \cite{Gro04a,Gro05a,Sre05a},
and it is unclear which one of them should be viewed as the chiral
partner of the yrast band. The yrast and B1 bands are of positive-parity
with known spins, while for band B3 the spin assignments are
tentative.

For all the $N=75$ isotones in question, PRM calculations were
performed \cite{Har01a,Sta01a,Sta02a,Pen03a}, and generally a good
agreement with experiment could be obtained for both the energies and
the B(M1)/B(E2) ratios, with a proper adjustment of the model parameters.
Lifetime measurements in $^{132}$La \cite{Gro04a,Gro05a,Sre05a}
revealed, however, that absolute values of the reduced transition
probabilities significantly deviate from these earlier PRM predictions.

The phenomenological TAC calculations were carried out for all the
considered nuclei as well \cite{Dim00a,Sta01b,Hec01a}. The
quadrupole deformations $\epsilon$ of $0.16$, $0.175$, $0.175$,
$0.195$ and triaxialities $\gamma$ of $39^\circ$, $32^\circ$,
$27^\circ$, $\approx26^\circ$ were found, respectively for
$^{130}$Cs, $^{132}$La, $^{134}$Pr, $^{136}$Pm. Chiral solutions were
obtained in a limited range of angular frequency. The lower limits
corresponded to the critical frequency, $\omega_{\text{crit}}$, discussed in
the present work. Only Refs.\ \cite{Dim00a} and \cite{Hec01a} quote their
values, which are $\omega_{\text{crit}}=0.3\,\mathrm{MeV}/\hbar$ for
$^{134}$Pr and $\omega_{\text{crit}}<0.2\,\mathrm{MeV}/\hbar$ for
$^{136}$Pm. Comparison with experimental energies was given only for
$^{134}$Pr in \cite{Sta01b,Dim00a}, where the behavior of the bands was
reproduced, on the average.


\subsection{Details of the calculations}
\label{detail_sub}

The present study is limited to the simplest chiral s.p.\
configuration,  $\pi{}h_{11/2}^1~\nu{}h_{11/2}^{-1}$, which has been
assigned in the literature to the candidate chiral bands observed in
the $N=75$ isotones. In order to check whether the existence of the HF
chiral solutions is not a particular feature of one Skyrme parameter
set, all calculations were repeated with two parametrizations, SLy4
\cite{Cha97a} and SKM* \cite{Bar82a}. The parity was kept as a
conserved symmetry, and the pairing correlations were not included.

In forming the chiral geometry, orientations of several
angular-momentum vectors play a crucial role. Their behavior depends,
among others, on the nucleonic densities which are odd under time reversal, like
the current and spin densities. Therefore, taking those densities and
corresponding terms in the energy density functional into account
seems important for the microscopic description of the chiral rotation.
Self-consistent methods are best suited for such a task, and
investigating the role of the time-odd densities was one of our
priorities in the present study.

The Skyrme energy density functional depends on time-even and time-odd nucleonic
densities with coupling constants $C_t^\rho$, $C_t^{\Delta\rho}$, $C_t^\tau$,
$C_t^J$, $C_t^{\nabla J}$ (10 time-even terms) and with coupling constants
$C_t^s$, $C_t^{\Delta s}$, $C_t^T$, $C_t^j$, $C_t^{\nabla j}$ (10 time-odd
terms) \cite{Dob95a}. The index $t=0,1$ denotes the isoscalar and isovector
parts. In standard parametrizations, which are used in the present work,
$C_t^\rho$ and $C_t^s$ additionally depend on the isoscalar particle density. If
the assumption of the local gauge invariance is made \cite{Dob95a}, there are
several following relations between the time-even and time-odd coupling
constants,
\begin{equation}
\label{eveodd_eqn}
C_t^j=-C_t^\tau~, \qquad C_t^J=-C_t^T~, \qquad C_t^{\nabla j}=+C_t^{\nabla J}~.
\end{equation}

In the present calculations, the coupling constants $C_t^J$ of the time-even
terms for $t=0,1$ were always set to zero, like in the original fits of the
forces SLy4 and SkM*. In order to conform to the local gauge invariance, whose
consequence is Eq.~(\ref{eveodd_eqn}), the coupling constants $C_t^T$ of the
time-odd terms were set to zero, too. Apart from $C_t^J$, all other time-even
coupling constants were taken as they come from the parameters of the Skyrme
force.

To examine the role of the time-odd densities, we have performed three
variants of calculations, throughout the text denoted as $N$, $G$, and $T$,
and defined in the following way:
\begin{itemize}
\item
Variant $N$: All time-odd coupling constants are set to zero and the
local gauge invariance conditions in Eq.~(\ref{eveodd_eqn}) are violated.

\item
Variant $G$: Coupling constants $C_t^j$ and $C_t^{\nabla j}$are taken
as they come from the parameters of the Skyrme force and as
required by the local gauge invariance (\ref{eveodd_eqn}); all
other time-odd coupling constants are set to zero.

\item
Variant $T$: Apart from $C_t^T$, all other time-odd coupling
constants are taken as they come from the parameters of the Skyrme
force and as required by the local gauge invariance (\ref{eveodd_eqn}).

\end{itemize}
The density-dependent and independent components of $C_t^s$ were
suppressed or not simultaneously. Setting or not some time-odd
coupling constants to zero implies excluding or including in the
calculations the corresponding time-odd terms of the mean field. In
variant $N$, the mean field contained only the time-even contributions
(apart from the cranking term) as in the case of the phenomenological mean
potentials.

The calculations were carried out by using the code HFODD (v2.05c)
\cite{Dob00a,Dob04a,Dob05a}. The code expands the s.p.\
wave-functions onto the HO basis, and uses the
iterative method to solve the HF equations. Twelve entire spherical HO shells
were included in the basis. It has been verified that increasing this
number up to 16, changes the quantities important for the present
considerations (deformation, moments of inertia, alignments etc.) by
less than 1\%.

\boldmath
\subsection{Energy minima in the $N=75$ isotones}
\unboldmath
\label{enemin_sub}

\newcommand{\myrule}  {\rule{0mm}{3.8mm}}
\renewcommand{\myrule}{\rule{0mm}{0mm}}
\begin{table*}[ht]
\caption{Quadrupole $\beta$ and $\gamma$ deformation parameters, parameters of the
classical model, $\cJ_{s,m,l}$ $[\hbar^2/\mbox{MeV}]$, $s_{s,l}$
$[\hbar]$, classical estimates for the critical frequencies and spins,
$\omega_{\text{crit}}^{\text{clas}}$ $[\mathrm{MeV}/\hbar]$ and $I_{\text{crit}}^{\text{clas}}$
$[\hbar]$, and the full HF TAC results for those quantities,
$\omega_{\text{crit}}^{\text{HF}}$ $[\mathrm{MeV}/\hbar]$ and $I_{\text{crit}}^{\text{HF}}$
$[\hbar]$, for the $N=75$ isotones.
The HF results with the SLy4 and
SkM* forces are shown for the $N$, $G$, and $T$ variants of calculation defined in Sec.\ \ref{detail_sub}.}
\label{crifre_tab}
\begin{center}
\begin{tabular}{l|lc|cc|ccc|cc|cccc}
\hline
 nucleus\myrule& \multicolumn{2}{c|}{force}
                      & $\beta$ & $\gamma$ & $\cJ_s$ & $\cJ_m$ & $\cJ_l$ & $s_s$ & $s_l$ & $\omega^{\text{clas}}_{\text{crit}}$ & $I^{\text{clas}}_{\text{crit}}$ & $\omega^{\text{HF}}_{\text{crit}}$ & $I^{\text{HF}}_{\text{crit}}$ \\
 \hline
$^{130}$Cs
&SLy4\myrule    & $N$ & 0.24    & 49       & 4.81    & 29.2    & 17.0    & 5.41  & 4.86  & 0.46                   & 12.8              &                      &                 \\
&               & $G$ & 0.24    & 49       & 5.50    & 37.1    & 21.3    & 5.45  & 5.16  & 0.37                   & 13.2              &                      &                 \\
&               & $T$ & 0.24    & 49       & 4.16    & 29.4    & 19.9    & 5.49  & 5.20  & 0.59                   & 16.8              &                      &                 \\
 \cline{2-14}
&SkM*\myrule    & $N$ & 0.23    & 47       & 5.86    & 31.3    & 17.9    & 5.43  & 5.01  & 0.43                   & 13.0              &                      &                 \\
&               & $G$ & 0.23    & 47       & 6.55    & 36.7    & 21.1    & 5.47  & 4.97  & 0.37                   & 13.0              &                      &                 \\
&               & $T$ & 0.23    & 47       & 5.69    & 33.4    & 20.0    & 5.49  & 5.14  & 0.43                   & 13.9              &                      &                 \\
 \hline
$^{132}$La
&SLy4\myrule    & $N$ & 0.26    & 46       & 7.18    & 28.7    & 19.1    & 5.44  & 4.90  & 0.57                   & 15.9              & 0.68                 & 18.8            \\
&               & $G$ & 0.26    & 46       & 8.45    & 36.0    & 23.7    & 5.60  & 5.21  & 0.47                   & 16.4              & 0.60                 & 20.3            \\
&               & $T$ & 0.26    & 46       & 7.12    & 31.7    & 22.2    & 5.64  & 5.26  & 0.60                   & 18.5              &                      &                 \\
 \cline{2-14}
&SkM*\myrule    & $N$ & 0.25    & 45       & 8.19    & 30.8    & 20.3    & 5.47  & 5.03  & 0.54                   & 16.0              & 0.62                 & 17.8            \\
&               & $G$ & 0.25    & 45       & 8.81    & 35.9    & 23.5    & 5.60  & 5.06  & 0.46                   & 15.9              & 0.54                 & 17.8            \\
&               & $T$ & 0.25    & 45       & 8.37    & 34.0    & 22.4    & 5.63  & 5.21  & 0.50                   & 16.5              & 0.58                 & 18.5            \\
 \hline
$^{134}$Pr
&SLy4\myrule    & $N$ & 0.26    & 58       & 6.11    &         & 25.4    & 5.00  & 4.36  &                        &                   &                      &                 \\
&{\it oblate}   & $G$ & 0.26    & 58       & 7.21    &         & 31.3    & 4.92  & 4.84  &                        &                   &                      &                 \\
&               & $T$ & 0.26    & 56       & 3.68    &         & 29.8    & 5.26  & 4.60  &                        &                   &                      &                 \\
 \cline{2-14}
&SkM*\myrule    & $N$ & 0.23    & 22       & 18.5    & 28.1    & 20.7    & 5.38  & 5.28  & 0.91                   & 25.0              &                      &                 \\
&{\it triaxial} & $G$ & 0.23    & 22       & 21.4    & 32.6    & 24.3    & 5.46  & 5.44  & 0.82                   & 26.1              &                      &                 \\
&               & $T$ & 0.23    & 22       & 20.7    & 30.8    & 24.8    & 5.52  & 5.57  & 1.08                   & 32.7              &                      &                 \\
 \hline
$^{136}$Pm
&SLy4\myrule    & $N$ & 0.25    & 53       &         &         & 24.7    &       & 4.62  &                        &                   &                      &                 \\
&{\it oblate}   & $G$ & 0.25    & 53       &         &         & 30.3    &       & 5.38  &                        &                   &                      &                 \\
&               & $T$ & 0.25    & 52       &         &         & 29.0    &       & 5.05  &                        &                   &                      &                 \\
 \cline{2-14}
&SkM*\myrule    & $N$ & 0.22    & 19       & 15.0    & 27.5    & 12.4    & 5.32  & 5.38  & 0.56                   & 14.8              &                      &                 \\
&{\it triaxial} & $G$ & 0.22    & 19       & 17.9    & 33.2    & 13.0    & 5.50  & 5.42  & 0.45                   & 14.4              &                      &                 \\
&               & $T$ & 0.22    & 19       & 17.4    & 29.8    & 13.4    & 5.51  & 5.53  & 0.56                   & 16.1              &                      &                 \\
 \hline
\end{tabular}
\end{center}
\end{table*}

As the first step, the HF calculations without cranking were performed to
find the $\pi{}h_{11/2}^1~\nu{}h_{11/2}^{-1}$ bandheads. Obviously,
the energetically most favored state of this configuration is
obtained if the valence proton particle and neutron hole occupy the
lowest and highest levels of the $h_{11/2}$ multiplets, respectively.
Table~\ref{crifre_tab} gives the obtained $\beta$ and $\gamma$
deformations for each isotone and Skyrme parameter set. They
practically do not depend on the included time-odd terms, and remain
almost constant when cranking is applied later on in our calculations. Note
that the present values of $\beta$ are up to 1.5 times larger than
those used in the previous phenomenological TAC calculations quoted
in Section~\ref{previous_sub}. Also the values of $\gamma$ are more distant from the maximum
triaxiality of $30^\circ$ as compared to the earlier results by other authors.

In $^{134}$Pr and $^{136}$Pm, two minima with the same $\pi h_{11/2}^1~\nu
h_{11/2}^{-1}$ configuration were found, which differ by the occupation of
positive-parity states. The energetically lower minima have similar
positive-parity s.p.\ structure as in $^{130}$Cs and $^{132}$La, but they
correspond to almost oblate shapes of $\gamma=53^{\circ}$-to-$58^{\circ}$. The
other minima have $\gamma=19^{\circ}$-to-$22^{\circ}$, thus corresponding to
triaxial shapes. In the following, those two kinds of minima in $^{134}$Pr and
$^{136}$Pm are conventionally referred to as {\it oblate} and {\it triaxial}.
Such a structure of minima and configurations appears for both interactions
studied here, SkM* and SLy4, and the corresponding sets of results are very similar
to one another. Therefore, to save space, below only the SkM* results are shown for
the {\it triaxial} minima, and only the SLy4 results for the {\it oblate} ones.

In Section~\ref{tac_sec}, we introduced the intrinsic frame of a
nucleus as formed by the principal axes of the quadrupole tensor.
Below, by the short ($s$), medium ($m$), and long ($l$) axes of our
triaxial solutions we understand the intrinsic axes indexed so that
$\langle x_s^2\rangle<\langle x_m^2\rangle<\langle x_l^2\rangle$,
where $x_i$ is the Cartesian coordinate for axis $i=s$, $m$, or $l$.


\boldmath
\subsection{Properties of the $h_{11/2}$ valence nucleons and of the core}
\unboldmath
\label{pac_sub}

We began our study of rotational properties by performing the standard
PAC calculations, in which we examined
rotations of the found, triaxial HF solutions about their short,
medium, and long axes. Let us recall that for the PAC about the principal
axis $i=s,m,l$, not only the cranking-frequency vector,
$\vec{\omega}$, but all the resulting s.p.\ and total mean
angular-momentum vectors, $\vec{j}$ and $\vec{I}$, have non-zero
components, $\omega_i$, $j_i$, $I_i$, only on that axis.

\begin{figure*}
\begin{center}
\includegraphics{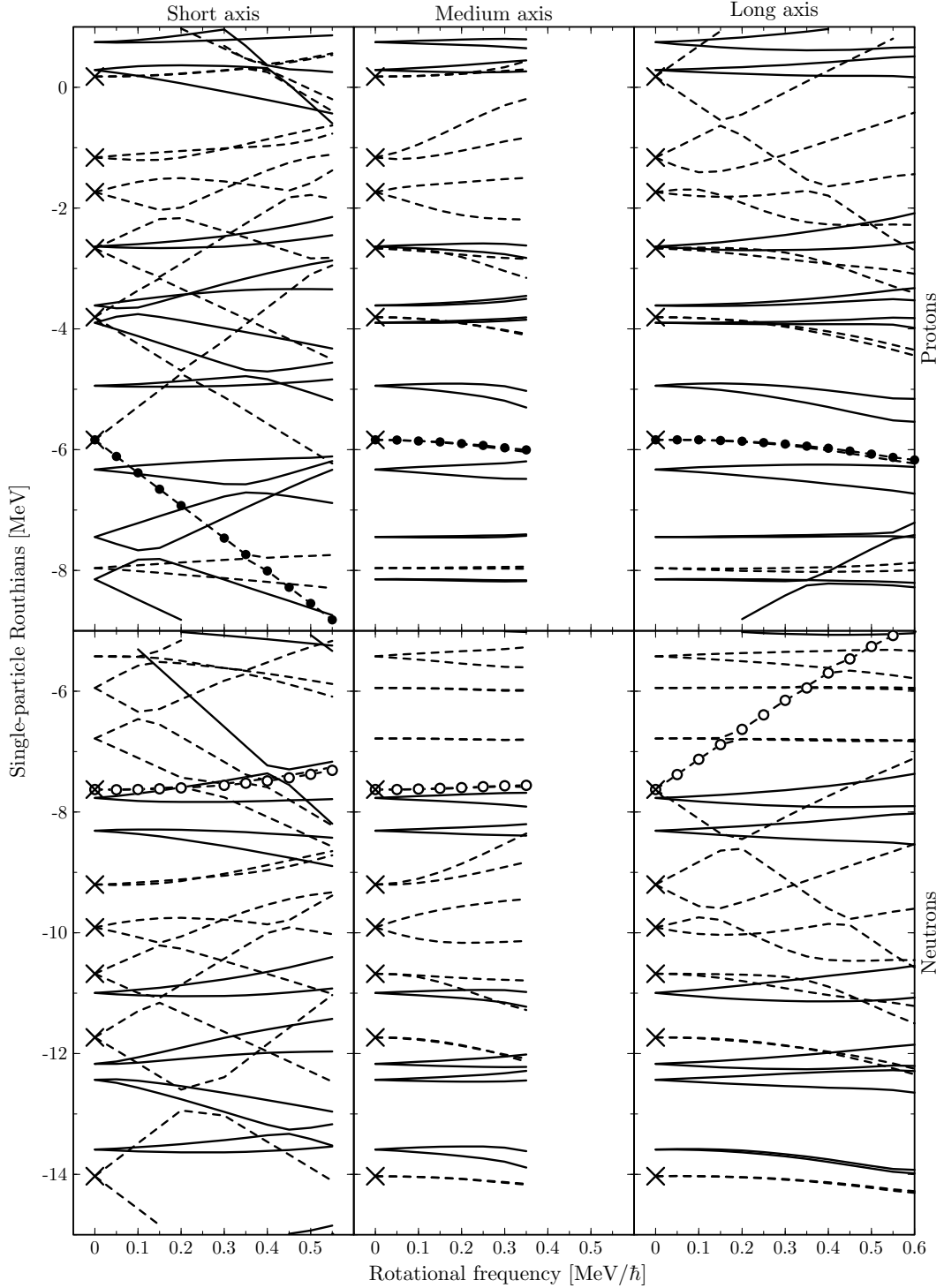}
\end{center}
\caption{Single-particle Routhians from the HF PAC calculations in $^{132}$La with
the SLy4 force and no time-odd fields. Proton (upper half) and neutron
(lower half) Routhians
for cranking about the short, medium, and long axes are shown.
Positive-parity and negative-parity levels are marked with solid and dashed
lines, respectively. Negative-parity levels belonging to the $h_{11/2}$ multiplet are
marked with crosses. The Routhians occupied by the valence $h_{11/2}$
proton particle and neutron hole are marked with full and open
circles, correspondingly.}
\label{pacrou_fig}
\end{figure*}

Figure~\ref{pacrou_fig} gives the s.p.\ Routhians obtained from the PAC
calculations in $^{132}$La with the SLy4 force, without the time-odd
fields. Routhians for other isotones, forces, and time-odd terms
included do not differ in their main features from the shown ones.
The upper and lower parts show the proton and neutron levels,
respectively, and the left, middle, and right panels correspond to
cranking about the short, medium, and long axes.
Positive-parity and negative-parity levels are marked with solid and
dashed lines. The negative-parity levels belonging to the $h_{11/2}$
multiplet are marked with crosses. The Routhians occupied by the valence
$h_{11/2}$ proton particle and neutron hole are marked with full and
open circles, respectively.

In both kinds of nucleons, the lowest levels of the $h_{11/2}$
multiplet split strongly for cranking about the short axis, and have
almost zero slope for the two remaining axes. The highest $h_{11/2}$
levels behave similarly, but split in the case of the rotation about the
long axis. The intermediate levels split for cranking about each
axis, but only weakly.

\begin{figure}
\begin{center}
\includegraphics{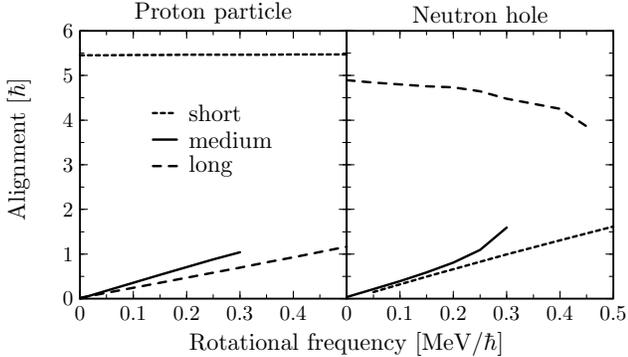}
\end{center}
\caption{Single-particle angular momentum alignments of the valence
$h_{11/2}$ proton particle and neutron hole in $^{132}$La, obtained
from the PAC about the short, medium and long axes. The HF results with
the SLy4 force are shown for the $G$ variant of calculation defined
in Sec.\ \ref{detail_sub}.}
\label{spalig_fig}
\end{figure}

For the PAC about the $i$-th axis, slopes of the s.p.\ Routhians, $e'$,
approximately translate
into the s.p.\ angular-momentum alignments, $j_i$, on that axis
according to the well-known formula
\begin{equation}
\label{jidedwi_eqn}
j_i \approx -\frac{\mathrm{d}e'}{\mathrm{d}\omega_i}~,
\end{equation}
which holds almost exactly in the present case, because
changes of the mean field with rotational frequency are negligible. Thus, one
can infer from Fig.~\ref{pacrou_fig} that the valence $h_{11/2}$
proton particle and neutron hole have non-zero alignments, $j^p_i$
and $j^h_i$, only for the PAC about short and long axes, respectively.
This can be seen directly in Fig.~\ref{spalig_fig}, which shows
$j^p_i$ and $j^h_i$ calculated as mean values of the angular-momentum
operator for the variant $G$ of the calculations with the SLy4 force.
Results for the short, medium, and long axes are plotted with dotted,
solid, and dashed lines.

It turns out that further important properties of the valence
$h_{11/2}$ particle and hole can be deduced from pure symmetry
considerations, which we present in Appendix~\ref{align_app}. We
consider an isolated two-fold degenerate s.p.\ level in a fixed
potential symmetric with respect to the $D_2$ group (rotations by
180$^\circ$ about the three principal axes). The main conclusion
relevant for the present case is the following. If in the PAC a s.p.\
state has a non-zero alignment, $j_i$, only for rotation about one
principal axis, then in the TAC its angular-momentum vector, $\vec{j}$,
will point along that axis and remain approximately constant regardless of the
length and direction of the applied cranking frequency,
$\vec{\omega}$. We call it a \emph{stiff alignment}. Thus, the angular
momenta, $\vec{j}^p$ and $\vec{j}^h$, of the valence $h_{11/2}$
proton particle and neutron hole are stiffly aligned along the short
and long intrinsic axes, respectively, as expected for the chiral
geometry.

In our self-consistent calculations, the s.p.\ spectrum of the
mean-field Hamiltonian, $\hat{h}$ of Eq.~(\ref{hhwj_eqn}), exhibits
the two-fold Kramers degeneracy only in variant $N$ of the
calculations, when no time-odd fields are taken. It seems that the
lowest and highest $h_{11/2}$ levels can be indeed treated as
isolated: their angular-momentum coupling to other s.p.\ states is
weak, which can be seen from the small curvature of their PAC
Routhians in Fig.~\ref{pacrou_fig}. The assumption about fixed
potential is also justified, because in our results the deformation
remains nearly constant with rotational frequency.

In the present case, the remnant coupling to other states and changes
of the mean field alter the ideal picture in that $\vec{j}^p$ and
$\vec{j}^h$ are not strictly constant, but show some remnant
dependence on the cranking frequency. As illustrated in
Fig.~\ref{spalig_fig}, this dependence is to
an approximation linear, and the PAC alignments can be written in the
form
\begin{equation}
\label{parhol_eqn}
\vec{j}^p\simeq s_s\vec{i}_s+\delta\cJ^p\vec{\omega}~, \qquad
\vec{j}^h\simeq s_l\vec{i}_l+\delta\cJ^h\vec{\omega}~,
\end{equation}
where $\vec{i}_s$ and $\vec{i}_l$ denote the unit vectors along the
short and long axes, respectively. The quantities $s_s$ and $s_l$ are
initial alignments at vanishing frequency, and $\delta\cJ^p$ and
$\delta\cJ^h$ are tensor coefficients, representing the s.p.\
contributions to the total inertia tensor, $\cJ$, of the nucleus.

The PAC allows to estimate the diagonal components of $\delta\cJ$,
which are slopes of the $j_i(\omega_i)$ curves plotted in
Fig.~\ref{spalig_fig}. One can see from the Figure that they may
attain up to $\sim$4\,$\hbar^2$/MeV. This is a significant value
compared to the total moments of inertia, discussed below and
collected in Table~\ref{crifre_tab}. For $^{123}$La, the total
moments are of the order of $\sim$8-36\,$\hbar^2$/MeV, depending on
the axis.

\begin{figure*}
\begin{center}
\includegraphics{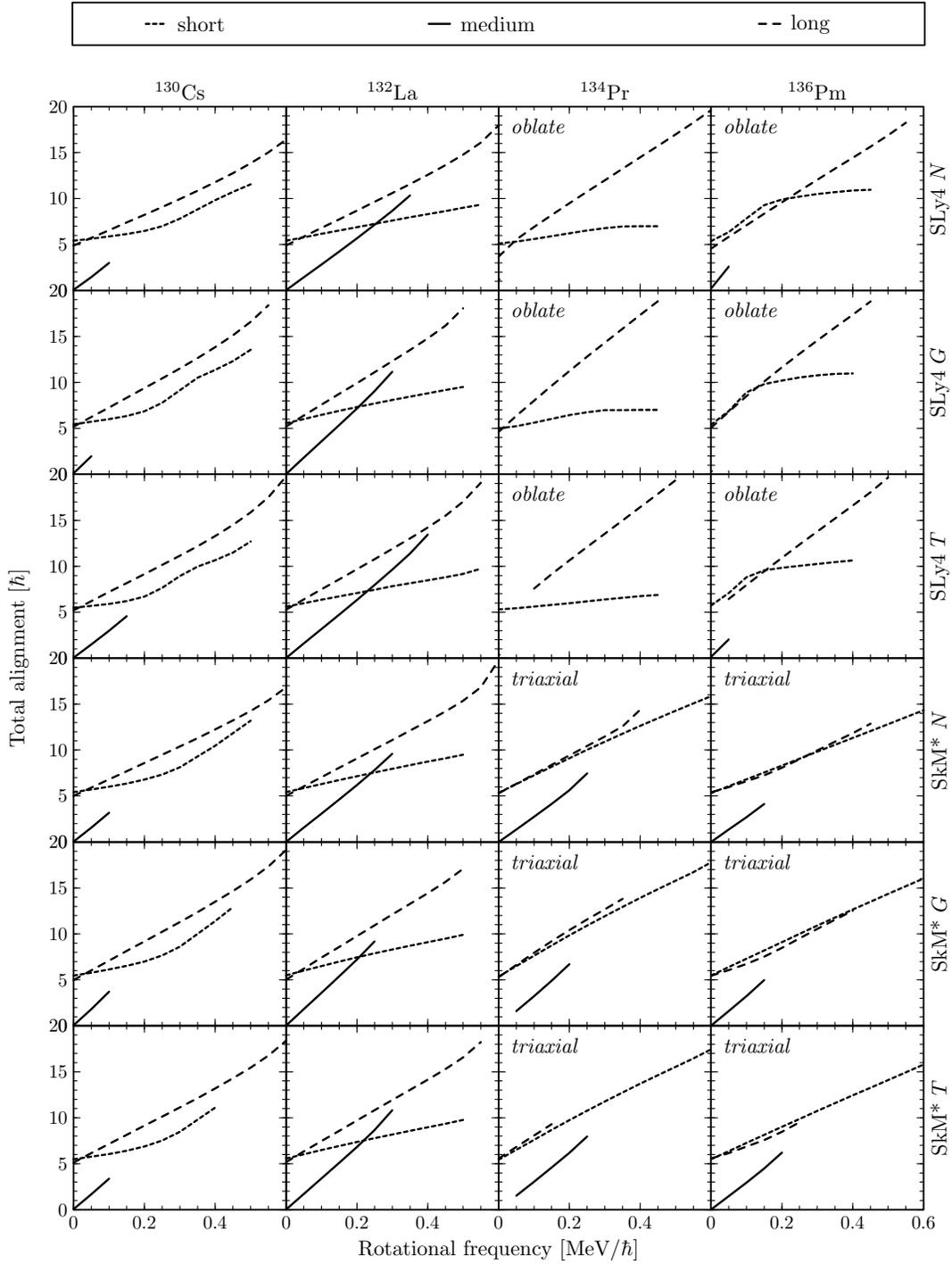}
\end{center}
\caption{Total angular momentum alignments obtained from the PAC about
the short, medium and long axes in $^{130}$Cs, $^{132}$La,
$^{134}$Pr, and $^{136}$Pm.
The HF results with the SLy4 and
SkM* forces are shown for the $N$, $G$, and $T$ variants of calculation defined in Sec.\ \ref{detail_sub}.}
\label{pacali_fig}
\end{figure*}

The same PAC calculations provide the total alignments, $I_s$, $I_m$, $I_l$, on
the short, medium, and long axes, respectively. They are plotted in
Fig.~\ref{pacali_fig} with dotted, solid, and dashed lines, correspondingly, for
all the considered cases. In the nearly {\it oblate} minima in $^{134}$Pr and
$^{136}$Pm, an attempt to crank around the medium axis leads to such a
self-consistent readjustment of the shape that solutions corresponding to the
rotation about the long axis are obtained.

The bends in the curves in
Fig.~\ref{pacali_fig}, like the ones for $I_s$ in $^{130}$Cs, are caused by
smooth crossings of the s.p.\ levels. Otherwise, the $I_i(\omega_i)$ dependence
is linear to a good approximation, and therefore the rotation can be called rigid.
The corresponding
slopes give the collective total moments of inertia, $\cJ_s$, $\cJ_m$, and
$\cJ_l$, with respect to the short, medium, and long axes. They contain the
valence particle and the valence hole  contributions, $\delta\cJ^p$ and
$\delta\cJ^h$, defined in Eq.~(\ref{parhol_eqn}).

At zero frequency, the
cranking around the medium axis gives a vanishing angular momentum, while the
cranking around the other two axes give non-zero limiting values, equal to the
initial alignments, $s_s$ and $s_l$, of the odd particle and hole. In view of
the considerations presented later on in this paper, we have extracted the
actual values of the parameters $\cJ_s$, $\cJ_m$, $\cJ_l$, and $s_s$, $s_l$, by
fitting straight lines to the calculated alignment vs.\ frequency curves shown in
Fig.~\ref{pacali_fig}. Whenever there was a bend in the calculated dependence,
the line was fitted in the rotational frequency range below the bend. Since the
alignment on the short axis for the {\it oblate} minimum in $^{136}$Pm shows a
particularly complicated behavior, we have not assigned any value to $\cJ_s$ in
this case. The obtained values of the parameters are listed in
Table~\ref{crifre_tab}, and discussed in Section~\ref{discus_sec}. These results
confirm that the moment of inertia with respect to the medium axis is the
largest.

The PAC method allows for estimating the diagonal components,
$\cJ_s$, $\cJ_m$, and $\cJ_l$, of the inertia tensor, $\cJ$. In order
to examine, as far as possible, the off-diagonal components we
performed a kind of perturbative test within the TAC method. We
applied to the non-rotating self-consistent solutions the cranking
frequency vector, $\vec{\omega}$, in several directions, performed
only one diagonalization of the resulting s.p.\ Routhian
(\ref{hhwj_eqn}), and investigated the response of the mean angular
momentum, $\vec{I}$. It turned out that the off-diagonal components
of $\cJ$ are negligibly small in all the considered isotones.

The microscopic results presented so far suggest that the considered
system can be modeled by two gyroscopes of spins $s_s$ and $s_l$
rigidly fixed along the short and long axes of a triaxial rigid rotor
characterized by the moments of inertia, $\cJ_s$, $\cJ_m$, $\cJ_l$,
of which $\cJ_m$ is the largest. It is instructive to solve the
associated problem of motion in the classical framework, which is
done in the next Section.

\subsection{Classical model}
\label{clasic_sub}

The classical model of chiral rotation was briefly introduced in Ref.\ \cite{Olb04a};
here we give its more detailed description and discussion.
In order to define the model, we
begin with elementary considerations related to the dynamics of rigid
bodies. By a classical gyroscope we understand an axial-shape rigid
body with moments of inertia, $\cJ_{\parallel}$ and $\cJ_{\perp}$,
with respect to the symmetry axis and an axis perpendicular to it,
respectively. Such a body spins with fixed angular frequency $\Omega$
around its symmetry axis; one can imagine that this motion is ensured
by a motor and frequency regulator such that $\Omega$ is strictly
constant in time.

Furthermore, let us imagine that the spinning body is
rigidly mounted on another rigid body that has triaxial inertia tensor and
three moments of inertia $\cJ'_s$, $\cJ'_m$, and $\cJ'_l$, with
respect to its short, medium, and long axis, respectively. Then, the
angular frequency $\Omega$ is maintained fixed with respect to the
triaxial body, irrespective of how the whole device moves, and the
angular frequency vector $\vec{\Omega}$ has by definition three
time-independent components $\Omega_s$, $\Omega_m$, and $\Omega_l$.
To simplify our considerations let us assume that the
axis of the gyroscope coincides with the short axis of the triaxial
body, i.e., $\Omega_s$=$\Omega$, $\Omega_m$=0, and $\Omega_l$=0.
In this case the principal axes of the tensor of inertia of the whole device
coincide with those of the triaxial body, and the three moments of inertia of
the device read,
\begin{equation}
\begin{array}{rcl}
\cJ_s &=& \cJ'_s + \cJ_{\parallel} , \\
\cJ_m &=& \cJ'_m + \cJ_{\perp}     , \\
\cJ_l &=& \cJ'_l + \cJ_{\perp}     . \\
\end{array}
\end{equation}

We assume that the device rotates with the total angular frequency
vector $\vec{\omega}$, which in the moving frame of the triaxial body
has components $\omega_s$, $\omega_m$, and $\omega_l$. In general,
these components may vary with time, although later we study only
such a motion of the device when they are time-independent. The kinetic
energy of the device is the sum of that of the triaxial body and gyroscope,
\begin{equation}
T=T_{\text{triax}}+T_{\text{gyro}},
\end{equation}
where
\begin{eqnarray}
T_{\text{triax}} &=& \tfrac{1}{2}\left(\cJ'_s\omega^2_s
                                      +\cJ'_m\omega^2_m
                                      +\cJ'_l\omega^2_l\right), \\
T_{\text{gyro}}  &=& \tfrac{1}{2}\left(\cJ_{\parallel}(\omega_s+\Omega)^2
                                      +\cJ_{\perp}\omega^2_m
                                      +\cJ_{\perp}\omega^2_l\right),
\end{eqnarray}
hence
\begin{equation}
\label{Ekin_eqn}
T                 =  \tfrac{1}{2}\left(\cJ_s\omega^2_s
                                      +\cJ_m\omega^2_m
                                      +\cJ_l\omega^2_l\right)
                     +  \cJ_{\parallel}\Omega\omega_s
                     +  \tfrac{1}{2}\cJ_{\parallel}\Omega^2.
\end{equation}
We see that the total kinetic energy is a sum of three terms. The
first one represents the rotation of the entire device irrespective
of the fact that it contains a spinning gyroscope; it depends only on
the total moments of inertia. The second one represents the
additional energy coming from the spinning gyroscope and depends on
its time-independent spin $s_s=\cJ_{\parallel}\Omega$, while the third one is
a constant which can be dropped from further considerations.

It is obvious that if we add two other gyroscopes aligned
with the medium and long axes and spinning with spins $s_m$ and
$s_l$, respectively, the second term can be simply written as a
scalar product $\vec{\omega}\cdot\vec{s}$, where vector $\vec{s}$ has
components $s_s$, $s_m$, and $s_l$. In this case, the total moment of
inertia $\cJ$ will be a sum of contributions from the triaxial body
and three gyroscopes. We note in passing that exactly the same
result is obtained for a single {\em spherical} gyroscope,
$\cJ_{\parallel}=\cJ_{\perp}$, tilted with respect to the principal
axes of the triaxial body in such a way that it has spin components
equal to $s_s$, $s_m$, and $s_l$.

The total angular momentum,
$\vec{I}$, of the system reads
\begin{equation}
\label{Spin_eqn}
\vec{I}=\cJ\vec{\omega}+\vec{s}~,
\end{equation}
where $\vec{s}$ is the above vector sum of spins of all the gyroscopes
and $\cJ\vec{\omega}$ stands for the tensor product of the moment of inertia
tensor $\cJ$ with the angular frequency vector $\vec{\omega}$.
In absence of potential interactions, the Lagrangian of the system is
equal to the total kinetic energy (\ref{Ekin_eqn}), generalized to the case of three gyroscopes, and thus is given by the formula
\begin{equation}
\label{LEkin_eqn}
L=\tfrac{1}{2}\vec{\omega}\cJ\vec{\omega}+\vec{\omega}\cdot\vec{s}~,
\end{equation}
where we dropped all constant terms.
Taking the laboratory components of $\vec{\omega}$ as generalized
velocities, it is easy to check that the generalized momenta are
equal to the laboratory components of $\vec{I}$. This fact allows us
to write the Legendre transformation \cite{Gol53a} and to obtain the
Hamiltonian of the system,
\begin{equation}
\label{HwIL_eqn}
H=\vec{\omega}\vec{I}-L=\tfrac{1}{2}\vec{\omega}\cJ\vec{\omega}~.
\end{equation}
Since the Lagrangian (\ref{LEkin_eqn}) does not depend explicitly on
time, the Hamilton function (\ref{HwIL_eqn}) is a constant of motion,
and is identified with the total energy, $E$, of the system. Consider
now a particular type of the Routhian \cite{Gol53a}, $H'$, namely
such that no variables undergo the Legendre transformation. In such a
case, $H'=-L$, and by rewriting the Routhian in terms of the
Hamiltonian one obtains
\begin{equation}
\label{HHwI_classic_eqn}
H'=H-\vec{\omega}\cdot\vec{I}~.
\end{equation}

Equations of motion for the model can be derived in the following
way. As for any vector, the time derivatives of the angular momentum vector,
$\partial_{t}\vec{I}$ -- taken in the
laboratory frame and $\partial^{\omega}_{t}\vec{I}$ -- taken
in a frame rotating with angular frequency
$\vec{\omega}$, are related by the formula
$\partial_{t}\vec{I}=\partial^{\omega}_{t}\vec{I}+\vec{\omega}\times\vec{I}$
\cite{Gol53a}. Since the angular momentum is
conserved in the laboratory frame, $\partial_{t}\vec{I}=0$,
one obtains the Euler equations \cite{Gol53a} for the time
evolution of the angular-momentum vector in the body-fixed frame,
\begin{equation}
\label{dwtI_eqn}
\partial^{\omega}_{t}\vec{I}=-\vec{\omega}\times\vec{I}~.
\end{equation}

The mean-field cranking approximation can only account for the
so-called \emph{uniform} rotations, in which the mean angular-momentum
vector is constant in the intrinsic frame of the nucleus,
$\partial^{\omega}_{t}\vec{I}=0$. Because of that, we restrict the
classical model studied here to such uniform rotations.
Due to Eq.~(\ref{Spin_eqn}), for uniform rotations also the
angular frequency vector is constant in the intrinsic frame.
The Euler equations (\ref{dwtI_eqn}) now take the form
$\vec{\omega}\times\vec{I}=0$, and require that $\vec{\omega}$ and
$\vec{I}$ be parallel. The same condition holds for the HF solutions
and is known as the Kerman-Onishi theorem; see Section~\ref{tac_sec}
and \cite{Ker81a}. The Euler equations can now be easily solved for the
considered classical model. However, to show further analogies with
the HF method, in what follows we find the uniform solutions by
employing a variational principle.

According to the Hamilton's principle, motion of a mechanical system
can be found by making the action integral, $\int L~dt$, stationary.
The real uniform rotations obviously belong to a wider class of trial
motions with $\vec{I}$ and $\vec{\omega}$ being constant in the intrinsic frame,
but not necessarily parallel to one another. Within
this class, Lagrangian (\ref{LEkin_eqn}) does not depend on time.
Therefore, extremizing the action for a given value of $\omega$=$|\vec{\omega}|$ reduces to
finding extrema of the Lagrangian as a function of the
intrinsic-frame components of $\vec{\omega}$. Since $H'=-L$, the
Routhian (\ref{HHwI_classic_eqn}) can be equally well used for this
purpose. This provides us with a bridge between the classical model and the
quantum cranking theory, where an analogous Routhian
(\ref{HHwJ_cranking_eqn}) is minimized in the space of the trial
wave-functions.

Extrema of $H'$ with respect to the intrinsic-frame components of
$\vec{\omega}$ at a given length of $\omega$ can be found by using a
Lagrange multiplier, $-\tfrac{1}{2}\mu$, for $\omega^{2}$
(the factor $-\tfrac{1}{2}$ being added for later convenience). We continue further
derivations for the case of two gyroscopes aligned along the $s$ and
$l$ axes, as dictated by the microscopic results presented in Section~\ref{pac_sub}, i.e., we employ the classical model for $s_m$=0.
Setting to zero the
derivatives of the quantity
\begin{eqnarray}
H'+\tfrac{1}{2}\mu\omega^2&=&\tfrac{1}{2}
\left[(\mu-\cJ_s)\omega_s^2+(\mu-\cJ_m)\omega_m^2\right.
\nonumber \\
&+&\left.(\mu-\cJ_l)\omega_l^2\right]-(\omega_s s_s+\omega_l s_l)
\end{eqnarray}
with respect to $\omega_{s}$, $\omega_{m}$, $\omega_{l}$, one obtains
\begin{eqnarray}
\omega_s            & = & s_s/(\mu-\cJ_s)~, \label{wsssmJs} \\
\omega_m(\mu-\cJ_m) & = & 0~,               \label{wmmJm0} \\
\omega_l            & = & s_l/(\mu-\cJ_l)~. \label{wlslmJl}
\end{eqnarray}
Equation~(\ref{wmmJm0}) gives either $\omega_{m}=0$ or
$\mu={\cJ}_{m}$, leading to two distinct classes of solutions.

\begin{figure*}
\begin{center}
\includegraphics{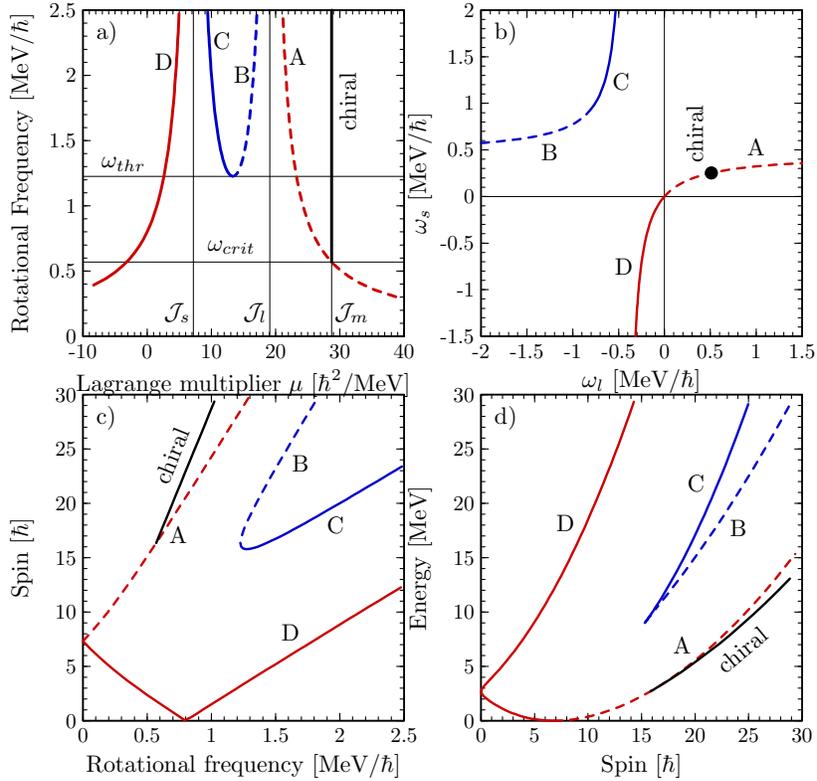}
\end{center}
\caption{(color online). Planar bands A, B, C, D, and the chiral band
obtained from the classical model. a) Rotational frequency,
$\omega(\mu)$. b) Intrinsic-frame trajectory of the rotational
frequency vector, $\vec{\omega}$. The chiral solution corresponds to
a straight line perpendicular to the figure plane, and intersecting
it at the marked point. That perpendicular direction represents
$\omega_m$. c) Angular momentum, $I(\omega)$. d) Energy, $E(I)$.}
\label{clasic_fig}
\end{figure*}

If $\omega_{m}=0$ then both $\vec{\omega}$ and $\vec{I}$ lie in the
$s$-$l$ plane. This gives planar solutions, for which the chiral
symmetry is not broken. All values of $\mu$ are allowed, and the
Lagrange multiplier must be determined from the length of
$\vec{\omega}$ calculated in the obvious way from (\ref{wsssmJs}) and
(\ref{wlslmJl}). Figure~\ref{clasic_fig}a shows $\omega$ versus $\mu$
for sample model parameters, extracted from the $^{132}$La HF PAC solutions with
the SLy4 force with no time-odd fields, and listed in
Table~\ref{crifre_tab}. The solutions marked as A and D exist for all
values of $\omega$, while above some threshold frequency,
$\omega_{\text{thr}}$, two more solutions appear, B and C. This threshold
frequency can be determined by finding the minimum of $\omega$ in
function of $\mu$, and reads
\begin{equation}
\omega_{\text{thr}}=\frac{\left(s_s^{2/3}+s_l^{2/3}\right)^{3/2}}{|\cJ_l-\cJ_s|}~.
\end{equation}
The value of $\omega_{\text{thr}}$ coming from the present HF calculations
is rather high, higher than $1\,\mathrm{MeV}/\hbar$. Since bands B
and C are situated far above the yrast line (see
Fig.~\ref{clasic_fig}d) they will not be subject of further analysis.

For $\mu$=$\cJ_m$, all values of $\omega_m$ are allowed, while
components in the $s$-$l$ plane are fixed at
$\omega_s=s_s/(\cJ_m-\cJ_s)$ and $\omega_l=s_l/(\cJ_m-\cJ_l)$.
Consequently, the angular momentum has non-zero components along all
three axes, and the chiral symmetry is broken. For each value of
$\omega$, there are two cases differing by the sign of $\omega_{m}$,
and thus giving the chiral doublet. The fact that $\omega_{s}$ and
$\omega_{l}$ are constant leads to the principal conclusion that
chiral solutions cannot exist for $\omega$ smaller than the critical
frequency
\begin{equation}
\label{crifre_eqn}
\omega_{\text{crit}}^{\text{clas}}=\left[\left(\frac{s_s}{\cJ_m-\cJ_s}\right)^2
                   +\left(\frac{s_l}{\cJ_m-\cJ_l}\right)^2\right]^{1/2}~.
\end{equation}
At that frequency, and with $\omega_m$=0, the chiral solution
coincides with the planar band A.

In the $\vec{\omega}$ space, the four planar solutions form a
hyperbola in the $s$-$l$ plane, while the chiral doublet corresponds
to a straight line perpendicular to that plane. These curves are
shown in Fig.~\ref{clasic_fig}b. Figure~\ref{clasic_fig}c gives the
angular momentum in function of rotational frequency for all the
presented bands. With increasing $\omega$, the so-called dynamical moment,
$\mathcal{J}^{(2)}\equiv\mathrm{d}I/\mathrm{d}\omega$, asymptotically approaches
$\cJ_l$ for bands A and B, and $\cJ_s$ for bands C and D. For the
chiral band, $I$ is exactly proportional to $\omega$ with the
coefficient $\cJ_m$. Thus, the critical spin, $I_{\text{crit}}^{\text{clas}}$,
corresponding to the critical frequency (\ref{crifre_eqn}) reads
\begin{equation}
\label{crispi_eqn}
I_{\text{crit}}^{\text{clas}}=\cJ_m\omega_{\text{crit}}^{\text{clas}}~.
\end{equation}

Figure~\ref{clasic_fig}d summarizes the energies in function of spin.
The spin quantum number, $I$, is related to the length, $I$, of the
angular momentum vector, $\vec{I}$, by the condition $I(I+1)=|\vec{I}|^2$. At
low angular momenta, the yrast line coincides with the planar band D.
Then it continues along the planar solution A. Since the moment of
inertia $\cJ_m$ is the largest, beyond the critical frequency the
chiral solution becomes yrast, thereby yielding good prospects for
experimental observation.

Altogether, the classical model described here is defined by five
parameters, $\cJ_s$, $\cJ_m$, $\cJ_l$, $s_s$, and $s_l$, which
are extracted form the microscopic HF PAC calculations and listed
in Table~\ref{crifre_tab}. The model can then be applied to predict
properties of the planar and chiral TAC bands, and these predictions
can be compared with the HF TAC results. Such a comparison is
presented and discussed in the following Sections.

\subsection{Planar solutions}
\label{planar_sub}

We began our self-consistent TAC calculations by finding planar
solutions corresponding to the classical band A. The first point of
each band of this kind was obtained by restarting the HF iterations from the
previously converged non-rotating solution and by applying the
initial cranking frequency vector with non-zero components on its
short and long intrinsic axes. Once convergence was achieved, the
obtained solution served in turn as the starting point for the next value
of the rotational frequency. We proceeded in this way with the
frequency step of 0.05\,MeV/$\hbar$. We followed each band
diabatically, i.e., by exciting particles near the Fermi level
whenever an empty and an occupied s.p.\ level of the same parity
were about to cross, so that always the states with the same physical
properties were occupied.

\begin{figure*}
\begin{center}
\includegraphics{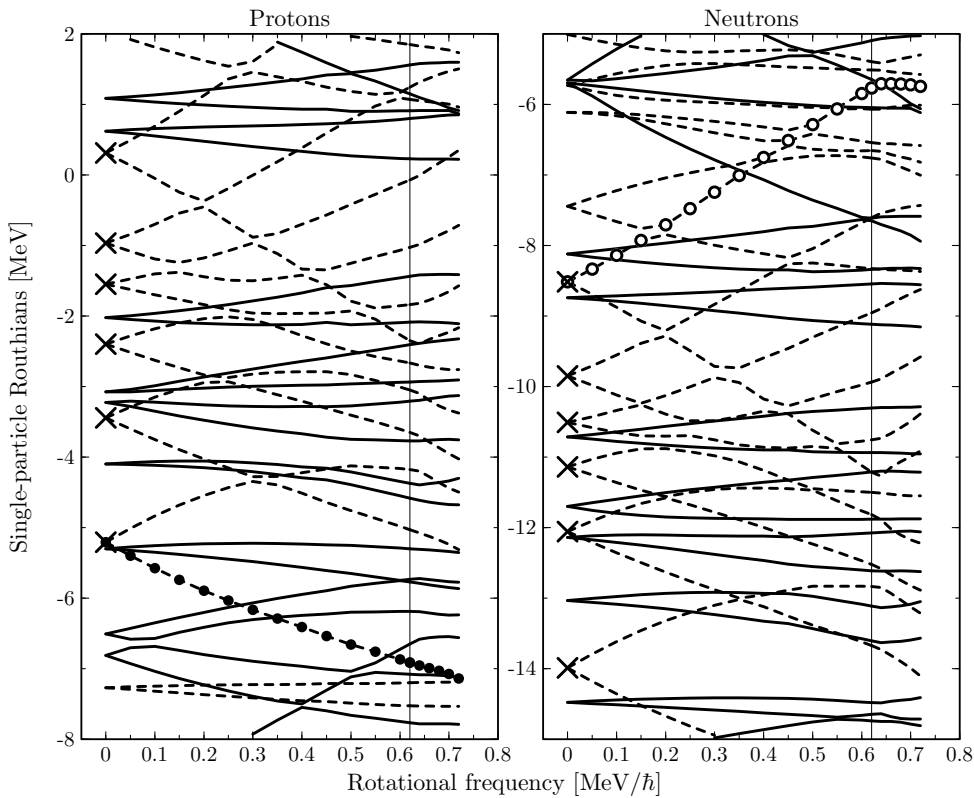}
\end{center}
\caption{Similar as in Fig.~\protect\ref{pacrou_fig} but for the
proton and neutron s.p.\ Routhians from the HF TAC calculations
with the SkM* force. The
thin vertical line is drawn at $\omega_{\text{crit}}^{\text{HF}}$. The Routhians to
the left and to the right of this line correspond to the planar and
chiral bands, respectively.}
\label{tacrou_fig}
\end{figure*}

In the solutions corresponding to a planar rotation, the cranking
frequency vector had non-zero components on the short and long
intrinsic axes in the self-consistent results. To give some insight
into the s.p.\ structure of those solutions, in Fig.~\ref{tacrou_fig}
we give the proton and neutron s.p.\ Routhians for the planar band in
$^{132}$La obtained with the SkM* force and no time-odd fields. Parts
of the plots to the left of the thin vertical lines are relevant for the planar results.
Contrary to the PAC Routhians of Fig.~\ref{pacrou_fig}, now both the
lowest and the highest $h_{11/2}$ levels split with rotational
frequency. This is consistent with the picture that the valence
particle and hole angular momenta, $\vec{j}^p$ and $\vec{j}^h$ ,
aligned on the short and long axes, respectively, now both have
non-zero projections on the tilted axis of rotation.

\begin{figure*}
\begin{center}
\includegraphics{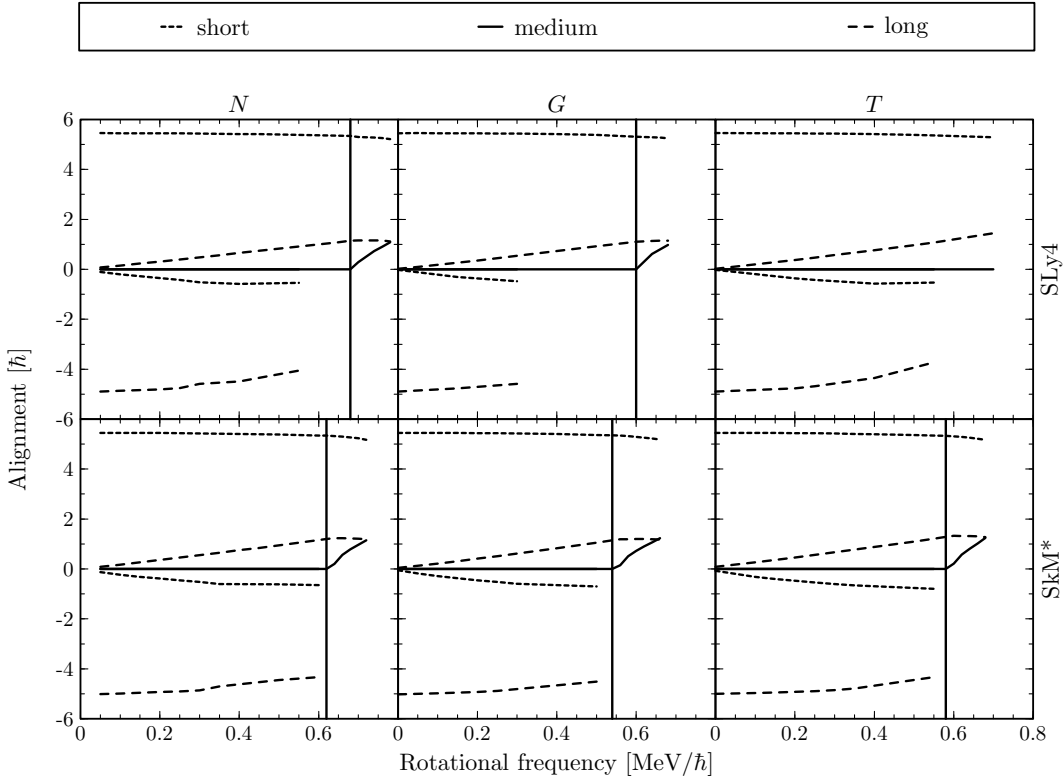}
\end{center}
\caption{Angular-momentum alignments of the lowest proton (positive
values) and highest neutron (negative values) $h_{11/2}$ levels on
the short, medium and long intrinsic axes from the HF TAC
calculations in $^{132}$La. The thin vertical line is drawn at
$\omega_{\text{crit}}^{\text{HF}}$. The curves to the left and to the right of this line
correspond to the planar and chiral bands, respectively.
The HF results with the SLy4 and
SkM* forces are shown for the $N$, $G$, and $T$ variants of calculation defined in Sec.\ \ref{detail_sub}.}
\label{tacali_fig}
\end{figure*}

In order to examine the angular momenta of the valence nucleons,
$\vec{j}^p$ and $\vec{j}^h$, in Fig.~\ref{tacali_fig} we plot their
projections onto the short (dotted line), medium (solid line) and long (dashed line)
intrinsic axes, for all the self-consistent
solutions in $^{132}$La. The positive and negative alignments
are those of the lowest proton and highest neutron $h_{11/2}$ levels;
the latter can be considered as representing $-\vec{j}^h$. Parts of
the plots to the left of the thin vertical lines concern the planar bands.
It can be seen that, indeed, the proton particle and the neutron hole
align their angular momenta on the short and long axes. Furthermore,
those alignments change rather weakly with rotational frequency which
means that the wave-functions are strongly confined by deformation
(deformation alignment).

\begin{figure*}
\begin{center}
\includegraphics{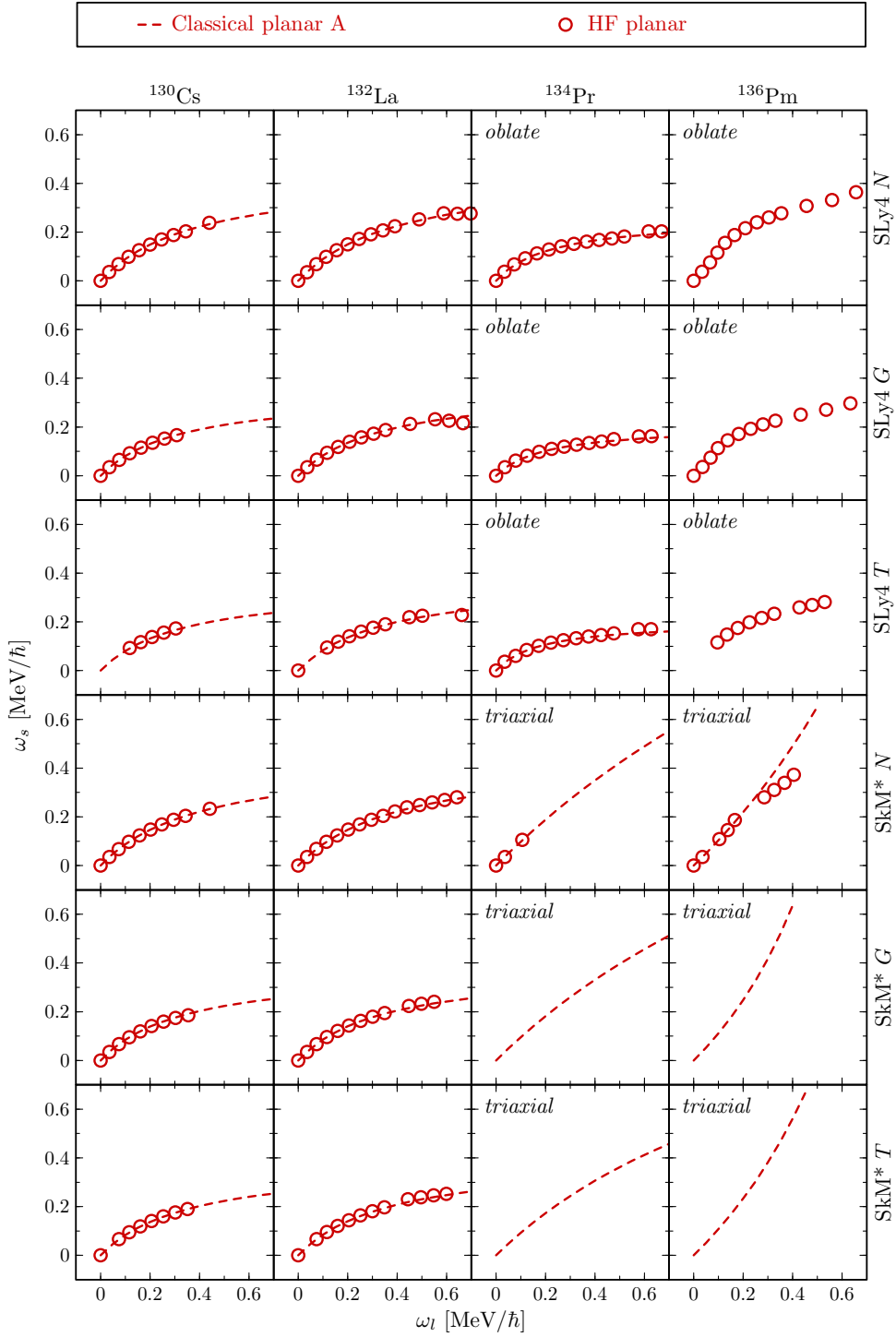}
\end{center}
\caption{(color online). Intrinsic-frame trajectories of the angular
frequency vector along the HF planar bands in the $N=75$ isotones,
compared to the classical solutions. The HF results with the SLy4 and
SkM* forces are shown for the $N$, $G$, and $T$ variants of calculation defined in Sec.\ \ref{detail_sub}.}
\label{plaome_fig}
\end{figure*}

It is worth emphasizing that the intrinsic-frame trajectories of
$\vec{\omega}$ along the self-consistent bands almost exactly follow
the classical ones, in all the considered cases. This is
illustrated in Fig.~\ref{plaome_fig}, where the dashed lines
represent classical A bands with parameters of
Table~\ref{crifre_tab}, and the HF results are marked with open
circles. For the {\it oblate} bands in $^{136}$Pm the classical line
is not shown because the parameters $\cJ_s$ and $s_s$ could not be
unambiguously extracted from the PAC calculations; see
Section~\ref{pac_sub}. However, the HF results follow a curve that
very much resembles the classical hyperbola. For the {\it triaxial}
minima in $^{134}$Pr and $^{136}$Pm, no self-consistent planar bands
could be obtained because of multiple level crossings.

Energy in function of spin also shows a striking agreement between
the classical and self-consistent results for the planar bands. This
can be traced in the case of $^{132}$La in Fig.~\ref{laener_fig},
by following the same symbols as those in Fig.~\ref{plaome_fig}. Some
deviations are visible only for rather high angular momenta.

\subsection{Chiral solutions}
\label{chiral_sub}

The planar HF solutions were easily obtained by applying small cranking
frequency increments to the non-rotating state. For chiral bands, analogous task
was more difficult, because these bands start at finite frequencies,
which in the present case are not lower than
$\approx0.5\,\mathrm{MeV}/\hbar$. Several level crossings may occur
between $\omega=0$ and such a high frequency, and it is difficult to
spot the required s.p.\ configuration at high frequencies. A
hint on how to  follow the $\pi h_{11/2}^1~\nu h_{11/2}^{-1}$
configuration diabatically comes from the classical prediction that
the chiral band branches off from the planar solution (at the point
corresponding to the critical frequency). One can thus restart
iterations from the planar band by applying cranking frequency with
non-zero component on the medium axis.

As the first step we performed a kind of perturbative search along the
planar bands, which turned out to be a very reliable test of where
the chiral solution may exist. Such a search
gives us also some understanding of why the chiral solutions do not appear in
several cases. The procedure we applied was the following.
To each converged point of the planar band, a small
additional component, $\omega_m$, of the angular frequency along the
medium axis was added, with $\omega_s$ and $\omega_l$ unchanged. The
resulting s.p.\ Routhian (\ref{hhwj_eqn}) was diagonalized only once.
Then, it was checked whether in the resulting non-selfconsistent
state the angular momentum and rotational frequency vectors were
parallel, as required by the Kerman-Onishi necessary condition of
self-consistency; see Section~\ref{tac_sec}. We can guess that in nuclei
stiff against deformation changes, the direction of $\vec{I}$ is the
only degree of freedom, and thus the Kerman-Onishi condition is also
sufficient. If $\vec{I}$ is parallel to $\vec{\omega}$ in the
non-selfconsistent state after one diagonalization, then it is very probable that
further iterations may lead to a converged chiral solution. Indeed,
this was always the case, and never a chiral solution was obtained,
in spite of several attempts, if that simple test gave negative
result.

\begin{figure*}
\begin{center}
\includegraphics{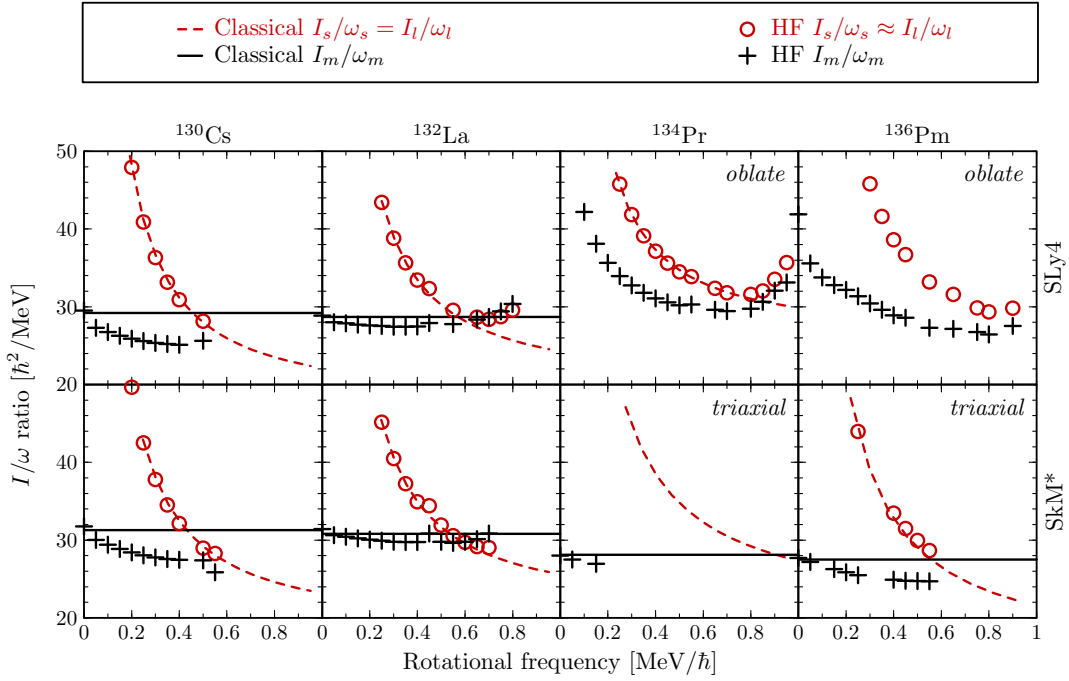}
\end{center}
\caption{(color online). Ratios
$I_m/\omega_m$ (open circles) and $I_s/\omega_s\approx I_l/\omega_l$ (plus symbols)
obtained in
the perturbative search for the HF chiral solutions along the planar
bands in the $N=75$ isotones; see text. Results for the SLy4 and SkM*
forces with no time-odd fields are shown.
Ratios $I_m/\omega_m$ (solid line) and $I_s/\omega_s=I_l/\omega_l$
(dashed line) corresponding to the classical-model chiral and planar
bands, respectively, are also plotted for comparison.}
\label{search_fig}
\end{figure*}

The condition for $\vec{I}$ and $\vec{\omega}$ being parallel can be
written in the form
\begin{equation}
\label{IsImIl_eqn}
\frac{I_s}{\omega_s}=\frac{I_m}{\omega_m}=\frac{I_l}{\omega_l}~.
\end{equation}
Note that the $I_s/\omega_s$ and $I_l/\omega_l$ ratios must be very
close to each other in the non-selfconsistent state, because the
Kerman-Onishi condition is fulfilled for the self-consistent planar
state. Therefore, the test consists in checking for each point of the
planar band if $I_m/\omega_m$ is equal to $I_s/\omega_s\approx
I_l/\omega_l$. In fact, this is reliable only if the time-odd fields
are switched off, because in their presence, the $I_m/\omega_m$ ratio
calculated perturbatively is significantly smaller than the
self-consistent result would be. The reason is that the relevant
components of the time-odd fields become active only after
self-consistency is achieved for a non-zero $\omega_m$. The test was
made with $\omega_m=0.05\,\mathrm{MeV}/\hbar$. The discussed ratios,
calculated for all the HF planar bands found in the $N=75$ isotones,
are plotted in Fig.~\ref{search_fig}. Plus symbols and
open circles denote the HF values of $I_m/\omega_m$ and
$I_s/\omega_s\approx I_l/\omega_l$, respectively.
To guide the eye, in the same Figure we also plotted the ratios
$I_m/\omega_m$ and $I_s/\omega_s=I_l/\omega_l$ corresponding to the
classical-model chiral and planar bands, respectively.

It can be a priori expected that chiral solutions do not appear in
the {\it oblate} minima in $^{134}$Pr and $^{136}$Pm, because of
insufficient triaxiality. Indeed, the calculated values of the
$I_m/\omega_m$ and $I_s/\omega_s\approx I_l/\omega_l$ ratios exhibit
a complicated behavior, and do not become equal to one another at any
point. In $^{130}$Cs, as well as in the {\it triaxial} minimum in
$^{136}$Pm, the two ratios clearly approach each other. It seems that
the only reason why they do not attain equality is that the planar
bands were not found up to sufficiently high frequencies, because of
level crossings. Note, however, that the moment of inertia associated
with the medium axis, $I_m/\omega_m$, significantly drops with
angular frequency, which takes $I_m/\omega_m$ away from
$I_s/\omega_s\approx I_l/\omega_l$, and defers their equalization to
higher frequencies. This effect is much weaker in $^{132}$La, where
the ratios do become equal, slightly above the point expected from
the classical model. Indeed, self-consistent chiral solutions were
found in this case, as described below.

After the first diagonalization of the perturbative test, the HF
iterations were continued in each case to achieve self-consistency.
In $^{130}$Cs, $^{134}$Pr, and $^{136}$Pm, as well as for low
rotational frequencies in $^{132}$La, the iterations converged back to
planar solutions. The same
result was obtained for different initial orientations of
$\vec{\omega}$ with respect to the intrinsic frame. This provides a
strong argument that, for the concerned configuration, no
self-consistent chiral solutions exist at low frequencies. In
$^{132}$La, for $\omega$ high enough, converged solutions were
obtained with $\vec{I}$ having non-zero components on all the three
intrinsic axes, which corresponds to chiral rotation. To examine the
chiral solutions independently of the planar ones, the found
fragments of chiral bands were used as starting points to obtain
solutions for lower and higher frequencies. Calculations were
performed with $\omega$ step of $0.02\,\mathrm{MeV}/\hbar$. At a
certain value of decreasing $\omega$, the planar orientation of
$\vec{I}$ was regained in the intrinsic frame, and the solution
merged into the previously found planar one. In a natural way, that
junction value of $\omega$ can be regarded as the Skyrme-HF result
for the critical frequency, and is denoted in the following as
$\omega_{\text{crit}}^{\text{HF}}$. Values of $\omega_{\text{crit}}^{\text{HF}}$ obtained in the
present calculations are collected in Table~\ref{crifre_tab}, and
discussed in Section~\ref{discus_sec}. On the side of highest
frequencies, chiral solutions were obtained up to a certain value of
$\omega$, and all attempts to go higher caused the iterations to fall
into a different minimum. This is probably due to multiple smooth
crossings of occupied and empty levels, particularly in neutrons.

\begin{figure*}
\begin{center}
\includegraphics{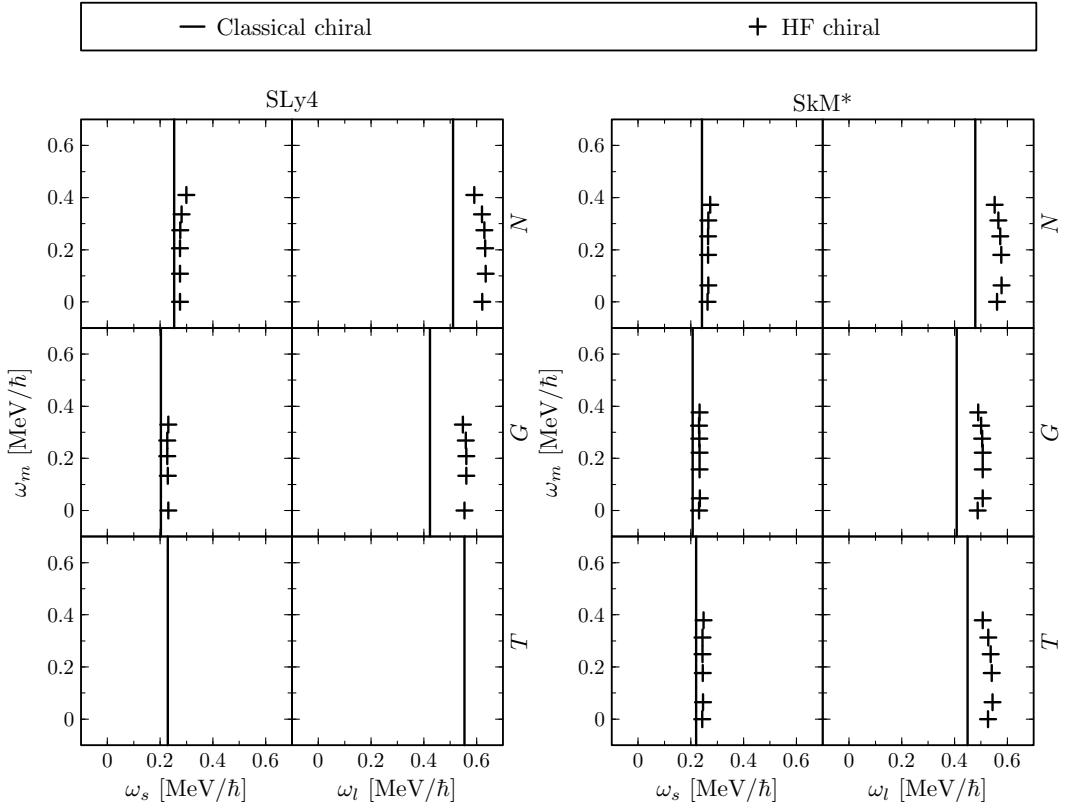}
\end{center}
\caption{Trajectories of the angular frequency vectors following the HF
chiral bands in $^{132}$La (plus symbols), compared to the classical solutions (solid lines).
Projections onto the $s$-$m$ and $l$-$m$ planes of the intrinsic
frame are shown. The HF results with the SLy4 and
SkM* forces are shown for the $N$, $G$, and $T$ variants of calculation defined in Sec.\ \ref{detail_sub}.}
\label{chiome_fig}
\end{figure*}

The HF results presented so far corroborate the main prediction of
the classical model, that chiral solution exists only above a certain
critical frequency, at which it branches off from the planar one. Also
the intrinsic-frame trajectory of $\vec{\omega}$ along the HF chiral
band is almost a straight line parallel to the medium axis, as in the
classical case. This is demonstrated in Fig.~\ref{chiome_fig}, which
shows projections of $\vec{\omega}$ on the $s$-$m$ and $l$-$m$
intrinsic planes for the HF (plus symbols) and classical (line) results.
The only difference is that $\omega_{\text{crit}}^{\text{HF}}$ is a bit higher than
$\omega_{\text{crit}}^{\text{clas}}$, and the HF line is shifted along the planar
band to higher frequencies.
Although the chiral solutions have been found in a rather narrow
$\omega$ interval, of about $0.1\,\mathrm{MeV}/\hbar$, the
accompanying increase in $\omega_m$ is significant, from zero to
about $0.4\,\mathrm{MeV}/\hbar$. This is so because $\omega_s$ and
$\omega_l$ are relatively large and almost constant along the chiral
solution.

Figure~\ref{tacrou_fig} shows the proton and neutron s.p.\ Routhians
for the chiral solution obtained in $^{132}$La with the SkM* force
and no time-odd fields. A thin vertical line is drawn at the value of
$\omega_{\text{crit}}^{\text{HF}}$. The Routhians to the left and to the right of
this line correspond to the planar and chiral bands, respectively.
Note first that the planar and chiral Routhians do indeed coincide at
$\omega_{\text{crit}}^{\text{HF}}$. The chiral Routhians do not seem to exhibit any
particular behavior. At high frequencies, the Routhian occupied by
the neutron hole enters into a region of high level density and the
corresponding s.p.\ state mixes with other negative-parity states.
Thus, it is doubtful whether the valence neutron hole can be identified with a single
state, and we do not examine its s.p.\ properties. The marking in
open circles is only tentative. However, the $h_{11/2}$ proton
particle is still well separated.

The alignments of the angular momentum, $\vec{j}^p$, of the
$h_{11/2}$ proton particle on the short, medium, and long intrinsic
axes for the SLy4 and SkM* forces and $N$, $G$, $T$ time-odd fields
are shown in Fig.~\ref{tacali_fig}. As in Fig.~\ref{tacrou_fig}, the
vertical line separates the planar and chiral bands. The plot
confirms the stiff character of those alignments in the chiral
solutions. Since the $\omega_s$ and $\omega_l$ components of the
cranking frequency vector hardly change along the chiral band, also
the considered alignments on those axes, $j^p_s$ and $j^p_l$, are
nearly constant. Only the projection on the medium axis, $j^p_m$,
increases due to the increase in $\omega_m$ from zero to about
$0.4\,\mathrm{MeV}/\hbar$.

\subsection{Separability of the TAC rotation}
\label{separa_sub}

To complete the presentation of our results, we here formulate and discuss
the separability rule, by which our planar (2D) and chiral (3D) HF
TAC solutions turn out to be simple superpositions of independent HF
PAC (1D) rotations about 2 or 3 principal axes, respectively.

\begin{figure*}
\begin{center}
\includegraphics{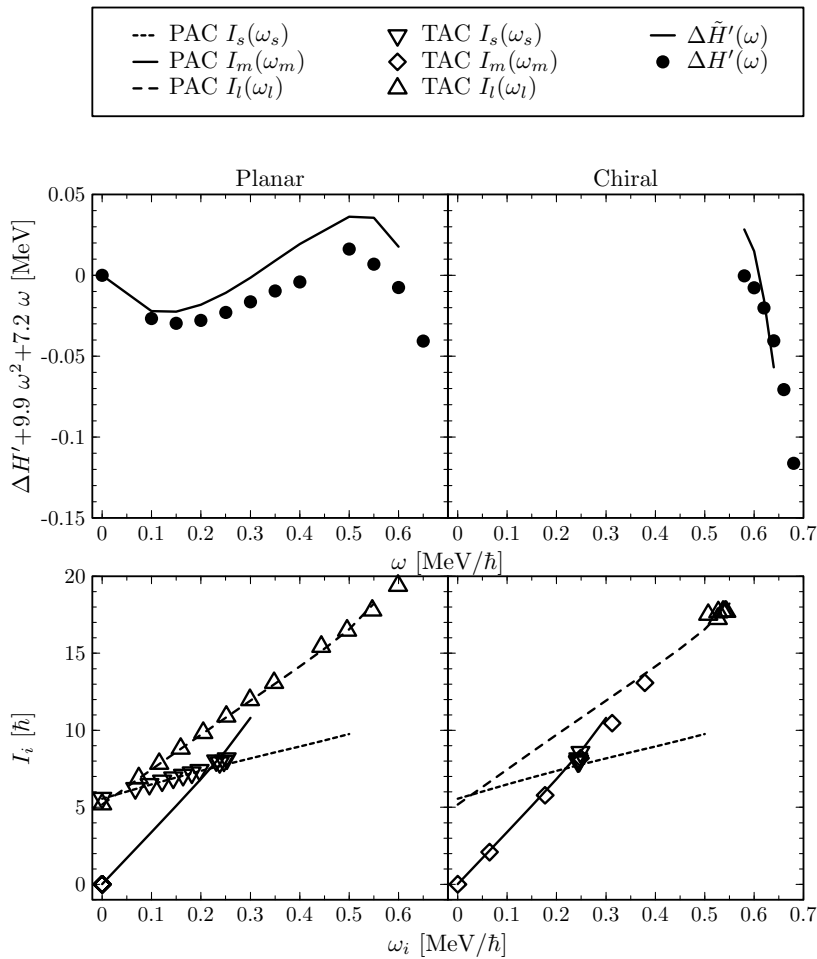}
\end{center}
\caption{Comparison of the full TAC (points) total angular-momentum
alignments (lower part) and Routhians (upper part) with analogous
quantities obtained within the PAC approach, as discussed in the text
(lines). The alignments $I_i$ on the principal axes $i=s,m,l$ are
shown as functions of the corresponding components, $\omega_i$, of
the angular frequency. The increments, $\Delta H'$
(\protect\ref{DeltaH}) and $\Delta\tilde{H}'$
(\protect\ref{DeltaHt}), in the full TAC and equivalent PAC Routhians
with respect to $\omega=0$ are plotted in function of the total
frequency $\omega$. In the latter plots, a quadratic function of
$\omega$ is added to stretch the scale. The HF results for the planar
(left) and chiral (right) solutions in $^{132}$La are shown for the
$T$ variant of the calculations (see Sec.\ \ref{detail_sub}) with the
SkM* force.}
\label{separa_fig}
\end{figure*}

Take, for instance, the total alignments on the principal axes,
$I_i(\omega_i)$, where $i=s,m,l$, as functions of the corresponding
components of the angular frequency. Such quantities were examined in
Section~\ref{pac_sub} for the PAC calculations, but they can be
equally well extracted for the planar and chiral solutions. A
comparison of the PAC and TAC results in $^{132}$La is shown in the
lower part of Fig.~\ref{separa_fig} for the $T$ variant of
calculations with the SkM* force. The PAC values of $I_s$, $I_m$, and
$I_l$ are plotted with lines, respectively: dotted, solid, and
dashed. The TAC results are marked with symbols, respectively:
down-triangles, diamonds, and up-triangles. The left and right parts
of the Figure show the TAC values of $I_i(\omega_i)$ for the planar
and chiral solutions, respectively. It is obvious that the results
obtained for the rotation about tilted axes are almost identical to
those obtained from the independent PAC calculations.

Unlike the angular momentum, the average value of the total TAC Routhian,
$\langle\hat{H}'\rangle$ of Eq.~(\ref{HHwJ_cranking_eqn}), cannot be
trivially decomposed into contributions from rotations about the three
principal axes. Yet, for each TAC solution, characterized
by an angular frequency $\omega$, we have its three intrinsic
components, $\omega_i$, and we can consider a sum, $\tilde{H}'$, of the
corresponding values of the PAC Routhians, $\langle\hat{H}'_i\rangle(\omega_i)$,
i.e.,
\begin{equation}
\label{equrth_eqn}
\tilde{H}'(\omega)  =  \sum_{i=s,m,l}\langle\hat{H}'_i\rangle(\omega_i)~,
\end{equation}
which we call \emph{equivalent Routhian}. We then compare the
difference relative to its value at $\omega$=0,
\begin{equation}
\label{DeltaHt}
\Delta\tilde{H}'(\omega)=\tilde{H}'(\omega)-\tilde{H}'(0),
\end{equation}
with the analogous difference,
\begin{equation}
\label{DeltaH}
\Delta H'(\omega)=\langle\hat{H}'\rangle(\omega)-\langle\hat{H}'\rangle(0),
\end{equation}
computed for the full TAC Routhian. In the upper part of
Fig.~\ref{separa_fig}, the differences (\ref{DeltaHt}) and
(\ref{DeltaH}) are plotted in function of $\omega$ as lines and
points, respectively.

For the chiral solution, as $\langle\hat{H}'\rangle(0)$ we take the
same value as for the planar case, because the chiral band can be
regarded as a continuation of the planar one, as discussed in
Section~\ref{chiral_sub}. The equivalent Routhian can be plotted only
in such a frequency range in which the PAC solutions are obtained for
the corresponding components $\omega_i$. In that range,
$\Delta\tilde{H}'$ deviates from $\Delta H'$ by
not more than 30\,keV, as it can be seen from the Figure, whereas the
total Routhian itself drops by about 8\,MeV between $\omega=0$ and
0.6\,MeV/$\hbar$. Therefore, we conclude that the equivalent
Routhian, constructed out of the PAC solutions, reproduces the full
TAC Routhian to a very high accuracy.

Another piece of information that is
contained in the TAC results, and not directly in the PAC results, is how the total
angular frequency, $\omega$, should be distributed over the three
principal axes for each given point of a band representing the rotation
about a tilted axis. That, however, can also be determined from the PAC
solutions by means of the Kerman-Onishi requirement \cite{Ker81a}
(see Section~\ref{tac_sec}) that the angular-momentum vector must be
parallel to the angular-frequency vector. Indeed, this condition
is equivalent to the system of three non-linear equations,
\begin{equation}
\begin{array}{rcl}
I_s(\omega_s)/\omega_s &=& \mu , \\
I_m(\omega_m)/\omega_m &=& \mu , \\
I_l(\omega_l)/\omega_l &=& \mu .
\end{array}
\end{equation}
By plotting the three PAC functions $I_i(\omega_i)/\omega_i$ on the
same plot as functions of $\omega_i$ we obtain a planar or chiral
solution whenever two or three, respectively, of these functions
cross a horizontal line. Then, values of $\omega_i$ corresponding to
the crossing points define the requested distribution of the
components within the total angular frequency $\omega$. The agreement
of the PAC and TAC alignments shown in Fig.~\ref{separa_fig}
guarantees that the above procedure gives a correct result.

We have thus demonstrated that within our HF results, the general 2D
or 3D rotation separates into two or three, in a sense independent, 1D
rotations about the principal axes. This is possible mainly because
our solutions are very stiff against deformation changes with
rotational frequency. Had they been soft, rotations about different
(principal or tilted) axes could cause different shape polarizations
that would prevent such a simple superposition of motions. Yet,
whenever the motions are separable, the PAC calculations supplemented
with the Kerman-Onishi condition are actually sufficient to describe
the 2D or 3D rotations. Although this rule has to be, in principle,
confirmed numerically in each particular case, it is plausible that
it will hold in all analogous cases of stiff alignments, by which the
difficult TAC calculations can be replaced by much more easily
performed PAC calculations.

\section{Discussion}
\label{discus_sec}

The main point left for the discussion concerns the values of the
critical frequency: where do they come from, whether they depend on the
Skyrme force used, how they are altered by the different time-odd
terms of the functional, how they would be influenced by the
inclusion of pairing, and what their relation to the experimentally
observed bands is. In this Section we give some remarks on these and
related topics.

Table~\ref{crifre_tab} summarizes the values of the classical-model
parameters, $\cJ_s$, $\cJ_m$, $\cJ_l$, $s_s$, and $s_l$, extracted
from the HF PAC results in Section~\ref{pac_sub}, of the classical
critical frequency, $\omega_{\text{crit}}^{\text{clas}}$, calculated from
Eq.~(\ref{crifre_eqn}), and of the critical frequency,
$\omega_{\text{crit}}^{\text{HF}}$, obtained from the HF TAC calculations as defined in
Section~\ref{chiral_sub}. In addition, the Table gives the
corresponding values of critical spins, $I_{\text{crit}}^{\text{clas}}$ and
$I_{\text{crit}}^{\text{HF}}$, respectively.

First of all, it can be seen from the examples of $^{130}$Cs and
$^{132}$La that the SLy4 and SkM* forces give quite similar values
for all the concerned quantities. The differences are not larger than
variations within one force due to taking into account different
time-odd fields. For all the four isotones in question, switching on
the fields $G$ increases all the moments of inertia, but particularly
$\cJ_m$, with respect to the case $N$. This causes a decrease in
$\omega_{\text{crit}}^{\text{clas}}$, but the corresponding $I_{\text{crit}}^{\text{clas}}$ does
not change much, because $\cJ_m$ is larger. Switching on the fields
$T$ results in values of $\cJ_m$ between those obtained for the cases
$N$ and $G$. The critical frequency always becomes higher than for
the $G$ fields, and the resulting critical spin is always the highest
among all the examined sets of time-odd fields. These variations in
$I_{\text{crit}}^{\text{clas}}$ are of the order of a few spin units.

In $^{132}$La, where the HF chiral solutions were found in
Section~\ref{chiral_sub}, values of the HF critical frequency and
spin, $\omega_{\text{crit}}^{\text{HF}}$ and $I_{\text{crit}}^{\text{HF}}$, are slightly higher
than the classical estimates, $\omega_{\text{crit}}^{\text{clas}}$ and
$I_{\text{crit}}^{\text{clas}}$. This can be understood on the basis of results of
the perturbative search for the HF chiral solutions, performed in
Section~\ref{chiral_sub}. As illustrated in Fig.~\ref{search_fig},
the perturbative ratio $I_m/\omega_m$, representing the moment of
inertia with respect to the medium axis, $\cJ_m$, slightly falls with
rotational frequency, which, according to Eq.~(\ref{crifre_eqn}),
causes the rise of the critical frequency. When the time-odd fields
are included, values of $\omega_{\text{crit}}^{\text{HF}}$ and $I_{\text{crit}}^{\text{HF}}$ vary
similarly to those of $\omega_{\text{crit}}^{\text{clas}}$ and $I_{\text{crit}}^{\text{clas}}$.

The HF method used in the present study does not take into account
the pair correlations. In order to include pairing in self-consistent
calculations one would have to apply the Hartree-Fock-Bogolyubov
(HFB) method in the TAC regime with two quasiparticle states blocked.
Present codes do not allow for such a study, and a systematic
investigation of pairing has to be left for future analyses. On the
one hand, one can expect that pairing effects may facilitate the
calculations by removing the sharp level crossings that occur in the
HF method, but on the other hand, convergence in presence of blocked
states may be more difficult to obtain. Supposing that the HFB TAC
results can be inferred from the HFB PAC calculations through a
determination of the parameters of the classical model, one could
gain some information on pairing effects by analyzing the HFB PAC
moments of inertia and alignments.

The critical frequency represents the transition point between planar
and chiral rotation. The notion of critical frequency was first
introduced in Ref.~\cite{Olb04a}, where the expression
(\ref{crifre_eqn}) for its value was derived from the simple
classical model presented in Section~\ref{clasic_sub}. However, the
occurrence of a transition from planar to chiral regime in the
structure of chiral bands was clear already from earlier
investigations. Both within the PRM and TAC, this effect was
obtained numerically but left without comment. This transition is abrupt in
the semi-classical cranking model, but may be rather smooth in real
nuclei. This is because the angular momentum vector oscillates about
the planar equilibrium below $\omega_{\rm crit}$, which corresponds
to non-uniform classical rotations \cite{Dim02a}, while above
$\omega_{\rm crit}$, it can still tunnel between the left and right
chiral minima, which represents chiral vibrations \cite{Sta01b}.

Because of these reasons, mean-field methods can provide quantitative
description of the bands in question only in the low-spin, planar regime,
where there is only one minimum. In the chiral region, the mean-field
approach does not take into account the interaction between the left
and right minima, which are exactly degenerate in energy, and the
experimentally observed energy splitting between the chiral partners
cannot be calculated. It is argued in the literature that the
mean-field chiral solution can be viewed as a kind of average of the
two partner bands, and thus mean trends can be compared. One can also
speculate about the value of the critical frequency or spin.
Description of the transition region is an interesting topic for
study invoking techniques beyond the mean field, like the
Generating-Coordinate Method.

Figure~\ref{laener_fig} gives a comparison of experimental and
calculated energies in $^{132}$La. Full symbols denote the
experimental yrast (circles), B1 (squares), and B3 (diamonds) bands,
discussed in Section~\ref{detail_sub}. Open circles and black
crosses represent the HF TAC planar and chiral solutions,
respectively. Their classical counterparts are marked with dashed
and solid lines. The HF results for the critical spin,
$I_{\text{crit}}^{\text{HF}}=15.9-18.5\hbar$, are rather high as compared to the
spin range, in which the bands B1 and B3 are observed. Yet, the
classical estimate, $I_{\text{crit}}^{\text{clas}}=9.2\hbar$, evaluated for the
Total-Routhian-Surface (TRS)
PAC results is already below that range \cite{Olb04a}. This means that the
inclusion of pairing in the calculations may be important for correct
interpretation of the data. However, from the closeness of
experimental spins to the possible values of $I_{\text{crit}}$ one can
suppose that, whichever of the bands B1 and B3 could eventually be interpreted as
the chiral partner of the yrast band, the concerned spin region may
actually represent the transition between planar and chiral rotation.
Although no HF chiral solutions were found in $^{130}$Cs, similar
conclusions can be drawn for that isotone on the basis of the
classical estimates of the critical spin. In case of $^{134}$Pr and
$^{136}$Pm it is not clear whether the {\it oblate} or {\it triaxial}
solutions should be taken for comparison with the bands observed in
those nuclei.

At low spins in $^{132}$La, where the supposed chiral partners have
not been observed, the yrast band is well reproduced by the HF planar
solutions, particularly with the time-odd fields included. This is
consistent with the supposed planar character of rotation at low
spins. Roughly at the spin where the chiral partners commence to be
visible, the yrast band significantly changes its behavior, which can
be attributed to entering into the chiral regime. In
this spin region the HF results agree semi-quantitatively with
experimental energies.

\section{Summary}

To conclude, the first Skyrme-Hartree-Fock (Skyrme-HF) calculations with
the Tilted-Axis Cranking were performed in $^{130}$Cs, $^{132}$La,
$^{134}$Pr, and $^{136}$Pm in the search for self-consistent solutions
corresponding to nuclear chiral rotation. Only the configuration
$\pi{}h_{11/2}^1~\nu{}h_{11/2}^{-1}$, earlier assigned to the
observed candidate chiral bands in those isotones, was considered.
Two Skyrme parametrizations, SLy4 and SkM*, were used. Terms
depending on time-odd nucleonic densities were either kept or
excluded from the Skyrme energy functional.

From the
Principal-Axis-Cranking analysis it was concluded that the
system in question can be modeled by two gyroscopes, representing the
valence particle and hole, with spins stiffly aligned with the short
and long axes of a triaxial rigid rotor, which stands for the core.
Such a model was analyzed in the classical framework. This led to
an important conclusion that chiral rotation can exist only above a
critical angular frequency, given by a simple expression.

The HF
solutions representing planar rotation were found in all the
considered $N=75$ isotones, and chiral solutions were obtained in
$^{132}$La. These solutions provide the first proof based on fully
self-consistent methods that nuclear rotation can attain a chiral
character. In all cases, the self-consistent solutions agree
surprisingly well with the results of the classical model, which
means that the model faithfully represents salient features of the
examined phenomenon.

It was found that the time-odd densities in the
energy functional have no qualitative influence on the results, and
change mainly the moments of inertia. The HF values of the critical
frequency are rather high as compared to the spin range in which the
candidate chiral bands were observed in $^{132}$La. The HF energies agree
satisfactorily with experiment only in the low-spin parts of the
bands, where the rotation is supposed to be planar. In the chiral
regime, the mean field is unable to reproduce the data
precisely, and the agreement is only qualitative. It seems, also,
that the experimentally observed bands actually represent a
transition from planar to chiral rotation.

The criteria used so far when attributing the chirality-partnership
interpretation to the experimentally found rotational bands that are based on
the 'small energy splitting' argument are clearly unsatisfactory on a long run.
Numerous superdeformed band studies (followed by the normal-deformation studies)
have shown that different intrinsic configurations and thus strictly speaking
different-shape nuclei may manifest nearly identical rotational bands.  It is
therefore necessary, to provide the experimental evidence going beyond just the
energy measurements, first of all the accompanying electromagnetic-transition
information. This could help excluding the mistake of interpreting e.g.\ the
shape coexistence phenomena in terms of chirality -- without providing extra
sufficient conditions. In this paper we were not able to propose any clear-cut
necessary-and-sufficient condition criteria to attribute the chirality label to
the experimental bands either. However, we do believe that the concept of the
critical frequency discussed in detail in this paper provides a useful tool in
interpreting the experimental results. The very fact that in some nuclei the
self-consistent HF calculations do provide chiral solutions is
highly non-trivial and very encouraging message in this field of research.

\acknowledgments
We would like to thank H. Flocard, J. Bartel, J. Stycze\'n, and W.
Satu\l{}a for valuable discussions. This work was supported in part
by the Polish Committee for Scientific Research (KBN) under Contract
No.~1~P03B~059~27, by the Foundation for Polish Science (FNP), and by
the French-Polish integrated actions program POLONIUM.

\begin{appendix}

\boldmath
\section{Rotational properties of single-particle states in a $D^T_2$-symmetric potential}
\unboldmath
\label{align_app}

In this Appendix, we discuss elementary rotational properties of
s.p.\ eigenstates of a mean-field Hamiltonian, which
is symmetric with respect to the $D^T_2$ group.
The group $D^T_2$ comprises the time-reversal operation, $\hat{T}$,
three signature operations, $\hat{R}_x$, $\hat{R}_y$, $\hat{R}_y$,
which are rotations through 180$^\circ$ about the three Cartesian
axes, and products of the time-reversal and signature operations,
which are called $T$-signatures, cf.\ Ref.~\cite{Dob00b} for more
information about this group in the context of mean-field
calculations.
In most cranking solutions corresponding to quadrupole
deformation, the group $D^T_2$ is a symmetry group of the s.p.\
Hamiltonian, $\hat{h}$, of Eq.~(\ref{hhwj_eqn}).
For a single Kramers
pair in a fixed potential, we investigate the response of the s.p.\
angular momenta to a cranking frequency applied in an arbitrary
direction. Our conclusions are based only on symmetry arguments, and
are thus independent on the particular implementation of the mean
field.

Irrespective of spatial symmetries, whenever the s.p.\
Hamiltonian is invariant under the time reversal, its spectrum
exhibits the two-fold Kramers degeneracy. We consider a single
Kramers pair, whose states are denoted as
$|\mu\rangle$ and $|\bar\mu\rangle$, where
\begin{equation}
\hat{T}|\mu\rangle=s_\mu|\bar\mu\rangle~, \qquad
\hat{T}|\bar\mu\rangle=s_{\bar\mu}|\mu\rangle~;
\end{equation}
$s_\mu$ is an arbitrary phase factor and
$s_{\bar\mu}=-s_\mu$.

All information about the matrix elements of the angular-momentum
operator, $\hat{\vec{J}}$, between the states $|\mu\rangle$ and
$|\bar\mu\rangle$ can be represented in a convenient way in terms of
the real \emph{alignment vector}, $\vec{J}^\mu$, and the complex
\emph{decoupling vector}, $\vec{D}^\mu$, of the state $|\mu\rangle$.
They are defined as
\begin{equation}
\vec{J}^\mu=\langle\mu|\hat{\vec{J}}|\mu\rangle~, \qquad \vec{D}^\mu=\langle\mu|\hat{\vec{J}}|\bar\mu\rangle~.
\end{equation}
Although the components of the decoupling vector change their phases when
$|\mu\rangle$ and $|\bar\mu\rangle$ change theirs, the relative phases of
those components do not depend on the phase convention. Since the
angular-momentum operator is odd under the time reversal, it can be easily
verified that
\begin{equation}
\vec{J}^{\bar\mu}=\langle\bar\mu|\hat{\vec{J}}|\bar\mu\rangle=-\vec{J}^\mu~, \quad \vec{D}^{\bar\mu}=\langle\bar\mu|\hat{\vec{J}}|\mu\rangle=\vec{D}^{\mu~*}~.
\end{equation}

If the s.p.\ Hamiltonian is symmetric with respect to the $D^T_2$
group, it is possible to chose the states forming the Kramers pair
as eigenstates of any of the three signature operators, $\hat{R}_i$, where
$i=x,y,z$, but only one at a time, because the signature operators do
not commute among themselves, i.e., for $i\neq j$
\begin{equation}
\hat{R}_i\hat{R}_j=\sum_{k=x,y,z}\epsilon_{ijk}\hat{R}_k.
\label{signatures}
\end{equation}
This results, respectively,  in three
formally different pairs, ($|\mu_i\rangle$,~$|\bar\mu_i\rangle$),
which are just three different bases in the same two-dimensional
eigenspace of $\hat{h}$.

We choose states
$|\mu_i\rangle$ so that they correspond to eigenvalues $-i$ under the action of
$\hat{R}_i$, while the eigenvalues of $|\bar\mu_i\rangle$ are $+i$.
Multiplication rules (\ref{signatures}) allow to easily express eigenstates
$|\mu_x\rangle$, $|\bar\mu_x\rangle$, $|\mu_y\rangle$, and $|\bar\mu_y\rangle$
through linear combinations of eigenstates $|\mu_z\rangle$ and $|\bar\mu_z\rangle$.
By fixing the relative phase
between states $|\mu_z\rangle$ and $|\bar\mu_z\rangle$
we obtain the following expressions:
\begin{eqnarray}
|\mu_x\rangle     & = &  \sqrt{\frac{-i}{2}}(|\mu_z\rangle+|\bar\mu_z\rangle)~, \label{mux_eqn} \\
|\bar\mu_x\rangle & = & -\sqrt{\frac{i}{2}}(|\mu_z\rangle-|\bar\mu_z\rangle)~,  \label{bmux_eqn}\\
|\mu_y\rangle     & = &  \sqrt{\frac{i}{2}}(|\mu_z\rangle+i|\bar\mu_z\rangle)~, \label{muy_eqn} \\
|\bar\mu_y\rangle & = &  \sqrt{\frac{-i}{2}}(i|\mu_z\rangle+|\bar\mu_z\rangle)~,\label{bmuy_eqn}
\end{eqnarray}
where $\sqrt{i}=\exp(i\pi/4)$ and $\sqrt{-i}=\exp(-i\pi/4)$. These
formulae allow to write $\vec{J}^{\mu_x}$, $\vec{D}^{\mu_x}$,
$\vec{J}^{\mu_y}$, and $\vec{D}^{\mu_y}$ in terms of $\vec{J}^{\mu_z}$ and
$\vec{D}^{\mu_z}$, i.e.,
\begin{eqnarray}
\vec{J}^{\mu_x} & = & \re \vec{D}^{\mu_z}~, \\
\vec{D}^{\mu_x} & = & -i\vec{J}^{\mu_z}-\im \vec{D}^{\mu_z}~, \\
\vec{J}^{\mu_y} & = & -\im \vec{D}^{\mu_z}~, \\
\vec{D}^{\mu_y} & = & \vec{J}^{\mu_z}-i\re \vec{D}^{\mu_z}~.
\end{eqnarray}

The fact that $|\mu_i\rangle$ and $|\bar\mu_i\rangle$ are eigenstates
of $\hat{R}_i$, together with the transformation rules of the
components, $\hat{J}_j$, of the angular momentum operator under the
three signatures,
\begin{equation}
\hat{R}_i^+\hat{J}_j\hat{R}_i=\left\{
\begin{array}{rcl}
+\hat{J}_j & \mbox{for} & j=i \\
-\hat{J}_j & \mbox{for} & j\neq i \\
\end{array}\right.~,
\end{equation}
induces limitations on the components, $J^{\mu_i}_j$ and $D^{\mu_i}_j$, of the alignment and decoupling vectors. Namely,
\begin{eqnarray}
J^{\mu_i}_j & = & \left\{
\begin{array}{rcl}
\mbox{non-zero} & \mbox{for} & j=i \\
0               & \mbox{for} & j\neq i \\
\end{array}\right.~, \label{alicon_eqn} \\
D^{\mu_i}_j & = & \left\{
\begin{array}{rcl}
0               & \mbox{for} & j=i \\
\mbox{non-zero} & \mbox{for} & j\neq i \\
\end{array}\right.~. \label{deccon_eqn}
\end{eqnarray}
In other words, $\vec{J}^{\mu_i}$ is confined to the axis $i$ and
$\vec{D}^{\mu_i}$ to the plane perpendicular to that axis.

From these relations and for the eigenstates defined as in Eqs.~(\ref{mux_eqn})--(\ref{bmuy_eqn}), it follows that all the quantities
$J^{\mu_i}_j$ and $D^{\mu_i}_j$ can be expressed through the three
``diagonal'' components, $J^{\mu_i}_i$, of the alignment vector,
\begin{eqnarray}
\vec{J}^{\mu_x}=(J^{\mu_x}_x,0,0)~, & & \vec{D}^{\mu_x}=(0,J^{\mu_y}_y,-iJ^{\mu_z}_z)~, \label{JD_mux_eqn} \\
\vec{J}^{\mu_y}=(0,J^{\mu_y}_y,0)~, & & \vec{D}^{\mu_y}=(-iJ^{\mu_x}_x,0,J^{\mu_z}_z)~, \label{JD_muy_eqn} \\
\vec{J}^{\mu_z}=(0,0,J^{\mu_z}_z)~, & & \vec{D}^{\mu_z}=(J^{\mu_x}_x,-iJ^{\mu_y}_y,0)~. \label{JD_muz_eqn}
\end{eqnarray}
Thus, all the information about the angular-momentum matrix elements within a
Kramers pair in the spectrum of a $D^T_2$-symmetric s.p.\ Hamiltonian
is contained in three real numbers, $J^{\mu_x}_x$, $J^{\mu_y}_y$, and $J^{\mu_z}_z$.

The symmetry group $D^T_2$ itself does not impose any conditions on
the``diagonal'' components, $J^{\mu_i}_i$. However, these
values can be further restricted if some other symmetry is present.
For example,
if $\hat{h}$ is axially symmetric, say with respect to the $z$ axis,
then the states $|\mu_z\rangle$, $|\bar\mu_z\rangle$ are eigenstates
of $\hat{J}_z$, which leads to quantization of $J^{\mu_z}_z$. In
fact, $J^{\mu_z}_z=+1/2,-3/2,...$, because
$\hat{R}_z=\exp(-i\pi\hat{J}_z)$, while in the adopted convention
$\hat{R}_z|\mu_z\rangle=-i|\mu_z\rangle$. For states $|\mu_x\rangle$
and $|\mu_y\rangle$, defined by (\ref{mux_eqn}) and (\ref{muy_eqn}),
one easily finds
\begin{eqnarray}
J^{\mu_x}_x & = & \frac{1}{2}\mathrm{Re}\langle\mu_z|\hat{J}_++\hat{J}_-|\bar\mu_z\rangle~, \label{Jxx_eqn} \\
J^{\mu_y}_y & = & \frac{1}{2}\mathrm{Re}\langle\mu_z|\hat{J}_+-\hat{J}_-|\bar\mu_z\rangle~, \label{Jyy_eqn}
\end{eqnarray}
where $\hat{J}_+=\hat{J}_x+i\hat{J}_y$ and
$\hat{J}_-=\hat{J}_x-i\hat{J}_y$ are the ladder operators, that
increment and decrement the magnetic quantum number, $J_z$, of an
eigenstate, $|J_z\rangle$, of $\hat{J}_z$,
\begin{equation}
\hat{J}_+|J_z\rangle\sim|J_z+1\rangle~, \qquad
\hat{J}_-|J_z\rangle\sim|J_z-1\rangle~.
\end{equation}
One can see, therefore, that the matrix elements in
(\ref{Jxx_eqn}) and (\ref{Jyy_eqn}) can be non-zero only if $|\mu_z\rangle$ and
$|\bar\mu_z\rangle$ differ in $J_z$ by one, that is if
$J^{\mu_z}_z=1/2$. In such a case,
$\langle\mu_z|\hat{J}_-|\bar\mu_z\rangle=0$, and
$J^{\mu_x}_x=J^{\mu_y}_y$. These results can be summarized as
\begin{equation}
\label{axi_eqn}
(J^{\mu_x}_x,J^{\mu_y}_y,J^{\mu_z}_z)=\left\{
\begin{array}{lll}
(J^{\mu_\perp}_\perp,J^{\mu_\perp}_\perp,J^{\mu_\parallel}_\parallel) & \mbox{for} & J^{\mu_\parallel}_\parallel=1/2 \\
(0,0,J^{\mu_\parallel}_\parallel)                                     & \mbox{for} & J^{\mu_\parallel}_\parallel=3/2,... \\
\end{array}\right.
\end{equation}
The parameter
\begin{equation}
\label{Jmpp_eqn}
J^{\mu_\perp}_\perp=\frac{1}{2}\mathrm{Re}\langle\mu_z|\hat{J}_+|\bar\mu_z\rangle
\end{equation}
is not restricted by the above kinematic conditions. It is related to the standard \emph{decoupling parameter}, $a=-2J^{\mu_\perp}_\perp$, considered by Bohr and Mottelson \cite{Boh75a}. They take such phases for the states $|\mu_z\rangle$ and $|\bar\mu_z\rangle$ that the $T$-signature-$y$, $\hat{R}_y^T=\hat{T}\hat{R}_y$, is the complex conjugation in the basis formed by these states. Since $\hat{J}_+$ is even under $\hat{R}_y^T$, in that convention the matrix element in Eq.~(\ref{Jmpp_eqn}) is real.

We now consider the TAC for a single Kramers pair
in a $D^T_2$-symmetric potential. We make two simplifying
assumptions. First, that the Hamiltonian $\hat{h}$ in the Routhian
$\hat{h}^\prime$ of Eq.~(\ref{hhwj_eqn}) does not change with rotational
frequency (non-selfconsistent cranking). Second, that the considered
Kramers pair has no coupling to other eigenstates of $\hat{h}$
through the angular-momentum operator (isolated pair). The TAC under such
conditions becomes a two-dimensional diagonalization problem, that
can be solved analytically.

We use the basis of states $|\mu_z\rangle$ and $|\bar\mu_z\rangle$.
For a degenerate Kramers pair, $\hat{h}$ reduces to its eigenvalue,
$e$. Matrix elements of the angular-momentum operator are defined by
Eq.~(\ref{JD_muz_eqn}). Altogether, matrix of the s.p.\ Routhian
(\ref{hhwj_eqn}) takes the form
\begin{eqnarray}
\label{roumat_eqn}
\hat{h}' & = & \hat{h}-\vec{\omega}\hat{\vec{J}} \\
         & = & \left[
\begin{array}{cc}
e & 0 \\
0 & e
\end{array}
\right]-\left[
\begin{array}{cc}
\omega_zJ^{\mu_z}_z                      &  \omega_xJ^{\mu_x}_x-i\omega_yJ^{\mu_y}_y \\
\omega_xJ^{\mu_x}_x+i\omega_yJ^{\mu_y}_y & -\omega_zJ^{\mu_z}_z
\end{array}
\right]~. \nonumber
\end{eqnarray}
It is easy to verify that the two eigenstates of this Routhian have
opposite-sign mean angular-momentum vectors $\vec{J}$, whose
components read
\begin{equation}
\label{alignm_eqn}
J_i=\pm\frac{\omega_i(J^{\mu_i}_i)^2}{\left(\omega_x^2(J^{\mu_x}_x)^2+\omega_y^2(J^{\mu_y}_y)^2+\omega_z^2(J^{\mu_z}_z)^2\right)^{1/2}}~.
\end{equation}
Equation~(\ref{alignm_eqn}) constitutes the central
point of discussion, because it defines the sought response of the
s.p.\ angular momenta to rotation under the assumed conditions. Note
that the dependence of $J_i$ on $\omega_i$ is non-linear in the general case.

Values of alignments (\ref{alignm_eqn}) are undefined if, and only
if, all the products $\omega_iJ^{\mu_i}_i$ vanish -- in particular
when $\omega=0$. In such a case, the Routhian (\ref{roumat_eqn}) is
proportional to unity, and the mean angular momenta of its
eigenstates depend on their (arbitrary) unitary mixing.

Two extreme cases of the dependence (\ref{alignm_eqn}) deserve
particular attention.
\begin{itemize}

\item
If $J^{\mu_x}_x=J^{\mu_y}_y=J^{\mu_z}_z=g$, then $J_i=g\omega_i/\omega$, and $\vec{J}$ always orients itself along $\vec{\omega}$, already for infinitesimal $\omega$. This can be dubbed \emph{soft alignment}.

\item
When only one of the parameters $J^{\mu_x}_x$, $J^{\mu_y}_y$, $J^{\mu_z}_z$ is non-zero, say $J^{\mu_j}_j$, then $J_i=J^{\mu_j}_j\delta_{ij}$, and $\vec{J}$ is independent of $\vec{\omega}$. We call this \emph{stiff alignment} on the $j$-th axis.

\end{itemize}

In axial nuclei, precisely one two-fold degenerate substate of each
deformation-split $j$-shell has $J^{\mu_\parallel}_\parallel=1/2$ and
$J^{\mu_\perp}_\perp\neq0$, which represents the soft alignment.
According to Eq.~(\ref{axi_eqn}), all other necessarily have a
vanishing decoupling parameter, and are thus stiffly aligned with the
symmetry axis. For prolate shapes, the lowest-energy substate has
$J^{\mu_\parallel}_\parallel=1/2$, and is soft, while for oblate
shapes it is the highest substate. In triaxial nuclei, values of the
parameters $J^{\mu_i}_i$, where $i=s,m,l$ corresponds to the short,
medium, and long principal axes, are equal to the s.p.\ alignments
obtained from the one-dimensional cranking about the three axes.
Indeed, for cranking about the axis $i$, the s.p.\ states are
eigenstates of $\hat{R}_i$. For example, from the results of
Section~\ref{pac_sub}, one can see that for the lowest $h_{11/2}$
substates of a triaxial nucleus only $J^{\mu_s}_s$ is non-zero, while
for the highest substates only $J^{\mu_l}_l$ does not vanish. These
alignments are thus stiff. Note that there are no states with stiff
alignment on the medium axis. The response to rotation of the middle
$h_{11/2}$ substates is soft, because all the three
parameters, $J^{\mu_s}_s$, $J^{\mu_m}_m$, $J^{\mu_l}_l$, are
non-zero.

In realistic cranking calculations the symmetry arguments discussed
here interplay with the fact that there is angular-momentum coupling
between different Kramers pairs and that the mean field does change
with rotational frequency. In the results of the present paper,
however, the change of deformation induced by rotation is negligible.
The angular-momentum coupling of the lowest and highest $h_{11/2}$
substates to other s.p.\ states is rather weak, what can be seen from
the small curvature of their one-dimensional Routhians in
Fig.~\ref{pacrou_fig}. The stiff character of their alignments is
fully confirmed in our self-consistent calculations, as discussed in
Section~\ref{pac_sub} and further in the paper. Investigation on how
suitable the notion of soft and stiff alignment is in other physical
cases remains a subject for further research.

\end{appendix}

\end{document}